\documentclass{article}

     \PassOptionsToPackage{numbers, compress}{natbib}

     \usepackage[preprint]{neurips_2021}

\usepackage[utf8]{inputenc} 
\usepackage[T1]{fontenc}    
\usepackage{hyperref}       
\usepackage{url}            
\usepackage{booktabs}       
\usepackage{amsfonts}       
\usepackage{nicefrac}       
\usepackage{microtype}      
\usepackage{xcolor}         

\usepackage{amsmath}
\usepackage{graphicx}
\usepackage{rotating}
\newtheorem{theorem}{Theorem}

\title{Mitigating Black-Box Adversarial Attacks via Output Noise Perturbation}

\author{%
  Manjushree B.~Aithal and Xiaohua~Li \\
  Department of Electrical and Computer Engineering \\
  Binghamton University, Binghamton, NY 13902, USA \\
  \texttt{\{maithal1, xli\}@binghamton.edu} \\
}

\begin{document}

\maketitle

\begin{abstract}
In black-box adversarial attacks, adversaries query the deep neural network (DNN), use the output to reconstruct gradients, and then optimize the adversarial inputs iteratively. In this paper, we study the method of adding white noise to the DNN output to mitigate such attacks, with a unique focus on the trade-off analysis of noise level and query cost. The attacker's query count (QC) is derived mathematically as a function of noise standard deviation. With this result, the defender can conveniently find the noise level needed to mitigate attacks for the desired security level specified by QC and limited DNN performance loss.  Our analysis shows that the added noise is drastically magnified by the small variation of DNN outputs, which makes the reconstructed gradient have an extremely low signal-to-noise ratio (SNR). Adding slight white noise with a standard deviation less than 0.01 is enough to increase QC by many orders of magnitude without introducing any noticeable classification accuracy reduction. Our experiments demonstrate that this method can effectively mitigate both soft-label and hard-label black-box attacks under realistic QC constraints. We also show that this method outperforms many other defense methods and is robust to the attacker's countermeasures.

\end{abstract}

\section{Introduction}

Along with the rapid development of deep neural networks (DNNs), there are a lot of online services, such as Clarifai API, Google Photos, advertisement detection and fake news filtering, etc., that highly rely on DNNs. Nevertheless, an intriguing issue is that DNNs are highly susceptible to small variations in input data \cite{szegedy2013intriguing}. Online DNN servers suffer from adversarial attacks where the attackers can slightly change the input data to make DNNs give false results or misclassification  \cite{papernot2017practical}. 

Depending on the knowledge about the DNNs that the attackers have, 
adversarial attacks can be classified into white-box attacks \cite{szegedy2013intriguing, goodfellow2014explaining,  carlini2017towards, madry2017towards} and black-box attacks \cite{chen2017zoo, tu2019autozoom, ilyas2018black, cheng2018query, cheng2019sign, alzantot2019genattack, guo2019simple, brendel2017decision}. The former assumes that the attackers have complete knowledge of the deep network, while the latter assumes that the attackers have limited knowledge, typically some output information of the DNNs. Compared with white-box attacks, black-box attacks are more realistic threats to real-world practical applications.

In general, black-box attacks need to estimate gradients via the output information of the deep networks obtained through querying and use these estimated gradients to optimize their adversarial inputs. The query cost is thus a critical constraint to attackers.
Over the recent years, more and more efficient black-box attack methods have been developed and they can now generate adversarial samples with only a few hundreds of queries \cite{tu2019autozoom, dong2020benchmarking}. Considering this fast increasing threat, it is the right time to develop effective defense methods \cite{guo2017countering, papernot2016distillation}. Unfortunately, most existing defense techniques are shown to provide a false sense of defense \cite{athalye2018obfuscated}.

In this paper, we study the performance of the simple output noise perturbation as a defense against black-box attacks, where the defender (or the DNN) adds white noise to the DNN outputs. Since it is impossible to find a technique that can completely stop attackers of unlimited resource, we focus on mathematical analysis of the attack-defense trade-off in terms of noise level and query count (QC). We believe such theoretical analysis is critical for defense study because it is computationally intractable to guarantee defense just with experiments. Specifically, we express QC as a function of noise standard deviation $\sigma$, with which the defender can easily apply appropriate noise to prevent attacks up to certain performance loss and security levels. For example, our results demonstrate that small noise with $\sigma \leq 0.01$ can prevent black-box attacks with a million query budget over the MNIST, CIFAR10 and IMAGENET datasets without noticeable classification accuracy loss. 


The major contributions of this paper are outlined as follows.
\begin{itemize}
\item We develop a new analysis framework to study the trade-off between noise level and QC mathematically instead of only heuristic approach via experiments. The signal-to-noise ratio (SNR) of the noisy gradients is derived, and it exhibits that small noise is magnified by the small DNN outputs. The attacker's QC is shown to be increased by many orders-of-magnitude with an extremely small noise.


\item We analyze the properties of the noise perturbation method and show that the proposed method is robust to the countermeasures of the attackers. We also observe that quantization and output-correlated noise does not perform well, which explains that output noise perturbation is better than other randomization or gradient obfuscation methods. 

\item We experiment with a list of representative black-box attack algorithms, including both soft-label and hard-label attacks. The results fit well with the analysis and demonstrate the effectiveness of this method against these attacks. 



\end{itemize}

This paper is organized as follows. Related works are presented in Section \ref{related}. The noise perturbation method is studied in Section \ref{analysis}.
Experiments are conducted in Section \ref{experiment}. Conclusions are given in Section \ref{conclusion}.

\section{Related work} \label{related}



Black-box attacks can be subdivided into three major classes: transfer-learning based attacks, soft-label attacks, and hard-label attacks \cite{ilyas2018black}. Transfer-learning based attacks exploit the fact that an adversarial input to one deep network may also be adversarial to another deep network \cite{papernot2017practical}. 

Soft-label attacks assume that the logit information is available to the attacker, either fully or partially. Narodytska et al. \cite{narodytska2016simple} used random perturbation and local searches to look for adversarial samples. Hayes et al. \cite{hayes2017machine} trained a generator neural network to generate adversarial samples. Chen et al. developed the zeroth-order optimization (ZOO)-based attacks \cite{chen2017zoo}, where they reconstructed gradients from output logits using zeroth-order gradient estimators. Ilyas et al. \cite{ilyas2018black} applied the natural evolution strategies (NES) to estimate the gradients. Tu et al. \cite{tu2019autozoom} improved the ZOO-based attacks with the AutoZOOM algorithm, which used autoencoders to generate gradient search directions. Cheng et al. \cite{cheng2019improving} combined the transfer-learning and ZOO-based attack techniques.

Hard-label attacks assume that only hard decisions of DNN outputs are available. Within this class, Brendel et al. \cite{brendel2017decision} exploited large perturbation to generate adversarial samples and used fine-tuning to reduce adversarial image distortion. Ilyas et al. \cite{ilyas2018black} picked a target image and fine-tuned it toward the original image. Cheng et al. \cite{cheng2018query, cheng2019sign} applied randomized gradient-free ZOO techniques.



On the defense side, a majority of existing studies are focused on white-box attacks. Most existing black-box defense techniques are in fact borrowed from their white-box version. A large number of  defense techniques were proposed based on the idea of gradient masking or gradient obfuscation, e.g., defensive distillation \cite{papernot2016distillation}, non-differentiable classifiers \cite{lu2017safetynet}, input randomization \cite{xie2018mitigating}, network structure randomization \cite{wang2019protecting}, etc. Unfortunately, almost all of them were defeated shortly after their publications via the so-called expectation-over-transformation (EOT) technique \cite{carlini2017adversarial, athalye2018robustness, athalye2018obfuscated}.
Today, the most effective way is perhaps adversarial training where adversarial samples are used to train the network \cite{tramer2017ensemble, samangouei2018defense, he2019parametric}, but the performance is not reliable for unknown attacks. 

To the best of our knowledge, the simple output noise perturbation method has not been studied in-depth. Dong et al. \cite{dong2020benchmarking} experimented with a long list of white-box/black-box attack/defense algorithms, but without this one. All the other reported noise perturbation techniques injected noise into the input or the network, not the output \cite{he2019parametric, liu2018towards, fan2019integration, li2018certified}. The reason is perhaps they were obtained from white-box attacks where adding noise to network outputs was of no use. Lee et al. \cite{lee2018defending} used output noise perturbation but for model stealing attack only and also without mathematical analysis.


\section{Analysis of Output Noise Perturbation} \label{analysis}

\subsection{Model of Black-Box Attack and Output Noise Perturbation Defense}

\begin{figure}[t]
	\centering

    \centerline{\includegraphics[width=0.5\linewidth]{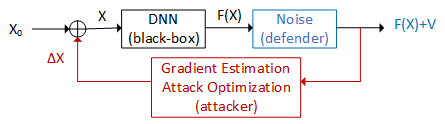}}
    
	\caption{The model of black-box attack and the defense via output noise perturbation.}
	\label{fig:model}
\end{figure}

Consider a DNN that classifies an image ${\bf x}_0$ into class $c$. The DNN outputs (softmax) logits $F({\bf x}_0)$, where ${F}$ is the DNN's nonlinear mapping function.  The classification result is $c=\arg\max_{i} {F}_i({\bf x}_0)$, where $F_i$ denotes the $i$th element function of $F$.

The objective of the adversarial attacker is to generate an image ${\bf x}={\bf x}_0+\Delta {\bf x}$ such that the DNN classifies it as $t = \arg\max_{i} F_i({\bf x}) \neq c$. Another aim of the attacker is that the difference $\Delta {\bf x}$ should be as small as possible. For the black-box attacks, the attackers query the DNN to obtain the input-output pair $({\bf x}, {F}({\bf x}))$, as shown in  Fig. \ref{fig:model}, with which they can minimize the following loss function to search for the adversarial sample ${\bf x}$ \cite{tu2019autozoom},
\begin{equation}
    f({\bf x}) = {\cal D}({\bf x}, {\bf x}_0) + \lambda {\cal L}(F({\bf x}), t),   \label{eq1.10}
\end{equation}
where ${\cal D}(\cdot, \cdot)$ is a distance function and ${\cal L}(\cdot, t)$ is the loss function. Typical distance functions are norms $\| {\bf x}-{\bf x}_0 \|_p$. Typical loss functions include the cross-entropy \cite{ilyas2018black} and the C\&W loss \cite{carlini2017adversarial}.

In this paper, we consider that the DNN defends itself by adding noise ${\bf v}$ to the logit and providing outputs $F({\bf x})+{\bf v}$. As a result, the attacker observes $F({\bf x})+{\bf v}$ instead of $F({\bf x})$. We assume that ${\bf v}$ is an independent and identically distributed (i.i.d.) Gaussian random vector with zero mean and standard deviation $\sigma$, i.e., ${\bf v}\sim {\cal N}({\bf 0}, \sigma^2{\bf I})$, where ${\bf I}$ is an identity matrix. We also assume that the DNN satisfies $\| F({\bf x}) - F({\bf y})\| \leq L \| {\bf x}-{\bf y} \|$ with a local Lipschitz constant $L$.   We consider low magnitude/small noise throughout our analysis.

 \noindent{\bf Definition 1}. Small noise is defined as the noise ${\bf v}$ whose standard deviation $\sigma$ is small so that $\log(1+{\bf v}) \approx {\bf v}$ is valid almost surely.


The performance of defense can be measured by three metrics: \textbf{attack success rate (ASR)}, \textbf{query count (QC)}, and \textbf{input distortion}. Strong and robust defense makes the black-box attacks result to low ASR, high QC, and high distortion. 
In this paper, we derive QC as function of $\sigma$ in theoretical analysis and calculate ASR as function of $\sigma$ under a pre-set QC limit in experiments. 
To save space, detailed analysis is presented for the soft-label targeted attack with the NES method only \cite{ilyas2018black}. Extension to other attack methods, hard-label and untargeted attacks, is explained briefly later in Section \ref{generalapp} because of the similarity in methods.

\subsection{Output Noise Perturbation to Mitigate NES Targeted Attack}  \label{formulation}

In \cite{ilyas2018black}, the soft-label NES targeted attack towards class $t$ is conducted by minimizing the softmax cross-entropy loss function $f({\bf x})= -\log {F_t({\bf x})}$, where $F_t({\bf x})$ is the softmax value of the target class. We have skipped the distance term ${\cal D}({\bf x}, {\bf x}_0)$ from (\ref{eq1.10}) in order to consider the most challenging defense situation. For the attacker it is easier to find an adversarial sample without the distortion constraint. 
The NES algorithm uses gradient descent to minimize the loss function iteratively. In each iteration, with $J$ queries, it estimates the gradient as
\begin{equation}
     \bar{\bf g} = \frac{1}{J}\sum_{j=1}^{J/2} {\bf g}_j, \;\;\;\; 
      {\bf g}_j = 
    {\bf u}_j \frac{1}{\beta}\log \frac{F_t({\bf x}-\beta {\bf u}_j)}{F_t({\bf x}+\beta {\bf u}_j)} = a {\bf u}_j, \label{eq1.30}
\end{equation}
where ${\bf u}_j$ is the random direction tensor and $\beta$ is the search variance. In particular, ${\bf g}_j$ is expressed as the attacker-generated tensor ${\bf u}_j$ multiplying a deterministic scalar multiplication factor $a$.

\begin{theorem} 
Assume white Gaussian noise perturbation with ${\bf v}\sim {\cal N}({\bf 0}, \sigma^2 {\bf I})$. The gradient ${\bf g}_j=a{\bf u}_j$ becomes ${\bf g}_j = A{\bf u}_j$ where $A = a+ \frac{1}{\beta} \log Z$ with random variable $Z\sim {\cal N}(1, \sigma_Z^2)$, 
\begin{align}
 \sigma_Z^2 &=  \frac{\sigma^2}{F_t^2({\bf x}-\beta {\bf u}_j)} + \frac{\sigma^2}{F_t^2({\bf x}+\beta {\bf u}_j)}.   \label{eq1.45}
\end{align} 
With small noise Definition 1, we have $A \sim {\cal N}(a, \sigma_Z^2/\beta^2)$. 
\end{theorem}
 
The detailed proof is shown in Appendix \ref{Theorem1Proof}. To understand the degree that  ${\bf v}$ randomizes the estimated gradient, we can evaluate the signal-to-noise ratio (SNR) of $A$ defined as ${\rm SNR}=\frac{a^2}{E[|\beta^{-1}\log Z|^2]}$, where $E[.]$ denotes mathematical expectation. We also call it the SNR of the noisy gradient.

\noindent{\bf Lemma 1}. {\it Under small $\beta$ and small noise, the SNR of $A$ is }
\begin{equation}
    {\rm SNR} = \frac{\left[ F_t({\bf x}-\beta {\bf u}_j) - F_t({\bf x}+\beta {\bf u}_j) \right]^2 F_t^2({\bf x}-\beta {\bf u}_j)} {\sigma^2 \left[ F_t^2({\bf x}-\beta {\bf u}_j) + F_t^2({\bf x}+\beta {\bf u}_j) \right]} 
    \leq \frac{2L^2\beta^2}{\sigma^2}.   \label{eq1.100}
\end{equation}

The detailed proof is shown in Appendix \ref{SNR1proof}.
We can see that the SNR is very small because the output variation $\Delta F_t = |F_t({\bf x}-\beta {\bf u}_j) - F_t({\bf x}+\beta {\bf u}_j)|$ and $\beta$ are very small in practice. A very small SNR makes $A$ to have opposite sign as $a$ with high probability, which changes the gradient descent toward the wrong direction and thus prevents the attacker's optimization from converging.

To derive QC as a function of noise level $\sigma$, we consider the following alternative approach since it is difficult to find QC expressions for deep networks. Consider the iterative gradient-descent minimization of $f({\bf x})=1/2[F({\bf wx})-F({\bf wx}^*)]^2$, where ${\bf w}$ is the weight of the input DNN layer and ${\bf x}$ is the DNN input. We assume that the function $F({\bf wx})$ is monotone between the starting point ${\bf wx}_0$ and the optimal point ${\bf wx}^*$ because otherwise there is no guarantee of convergence.  The minimization is conducted as ${\bf x}_{n+1}={\bf x}_{n}-a \partial{f({\bf x}_n)}/\partial{{\bf x}_n}$, $n=0, 1, \cdots$. Our objective is to find the ratio $R$, i.e., the ratio of the iteration number needed when using a constant learning rate $a$ to that when using the random learning rate $A =a + \sqrt{{\rm SNR}} v$ with noise $v$. 

\begin{theorem} If the learning rate $a$ is small such that $(1-a\lambda)^n \approx 1-na\lambda$, then
\begin{equation}
    R =  
    \frac{1}{4}\left(\sqrt{K^2+4} - K \right)^2, \;\;\;\;
    K = \frac{\Phi^{-1}(\epsilon)\sqrt{a\lambda}} {\sqrt{{\rm SNR}(1-\eta/v_0)}},  \label{eq1.101}
\end{equation}
where $\eta$ and $\epsilon$ are small probabilities, $\lambda$ and $v_0$ are constants related to ${\bf w}$, ${\bf x}_0$ and ${\bf x}^*$, and $\Phi^{-1}(\epsilon)$ is the inverse of the standard normal cumulative distribution function.
\end{theorem}

The detailed proof is shown in Appendix \ref{Theorem2proof}. From the proof, we also see that $R$ can be used as an estimation of $QC(noise)/QC(noiseless)$, i.e., the ratio of QCs between the case with noise perturbation and the case without noise perturbation. 

The relationship between QC and noise $\sigma$ can be readily analyzed based on (\ref{eq1.100}) and (\ref{eq1.101}). In particular, if $\sigma$ is small, then $R \sim 1/{\rm SNR}$, i.e., increases with 1/SNR. As a rule of thumb, $\log(R)$, $-\log{\rm SNR}$, and $\log \sigma - \log \Delta F_t$ change linearly with each other.

\begin{figure}[t]
	\centering

    \centerline{\includegraphics[width=0.5\linewidth]{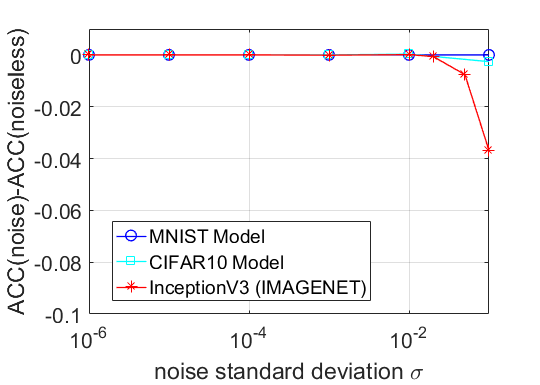}  \includegraphics[width=0.5\linewidth]{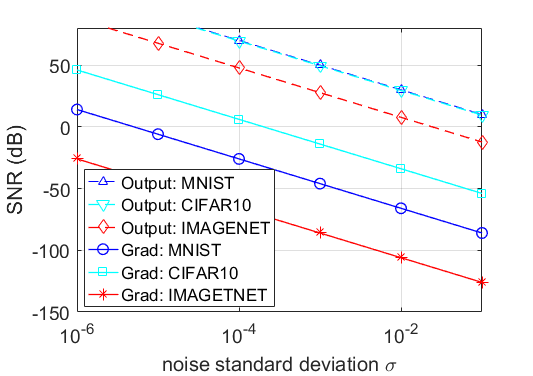}}
    
    \centerline{(a)  \hspace{0.5\linewidth} (b)} 
	\caption{(a) Classification accuracy degradation due to noise perturbation. (b) SNR of noisy model outputs $F({\bf x})+ {\bf v}$ ({\bf Output}) and SNR of gradient multiplication factor $A$ ({\bf Grad}).}
	\label{fig:snr}
\end{figure}

To understand better the tradeoff between performance loss (specified by $\sigma$) and defense security (specified by QC), we need to know the output variation $\Delta F_t$. For this we trained a 5-layer CNN model for the MNIST dataset, a 6-layer CNN model for the CIFAR10 dataset, and used the Inception-V3 model for the IMAGENET dataset. First, using their validation datasets, we calculated the statistical parameters of DNN outputs, which are shown in Table \ref{tbl:valparams}. We applied random ${\bf u}_j$ to calculate output variation $\Delta F_t$.  Second, we added noise to the outputs and calculated output SNR and ACC degradation. The results in Fig. \ref{fig:snr}(a) clearly show that there is almost no ACC degradation when noise $\sigma \leq 0.02$. From Fig. \ref{fig:snr}(b) we also find that the output SNR is high. 


Next, using the mean $\Delta F_t$ data in Table \ref{tbl:valparams}, we calculated the SNRs of $A$ and showed them in Fig. \ref{fig:snr}(b). The SNRs drastically reduced to very small numbers. At $\sigma=0.01$, the SNRs were -66dB, -34dB and -106 dB for the MNIST, CIFAR10 and IMAGENET models. Such low SNRs made $A$ completely different from $a$. 

\begin{figure}[t]
	\centering

    \centerline{\includegraphics[width=0.5\linewidth]{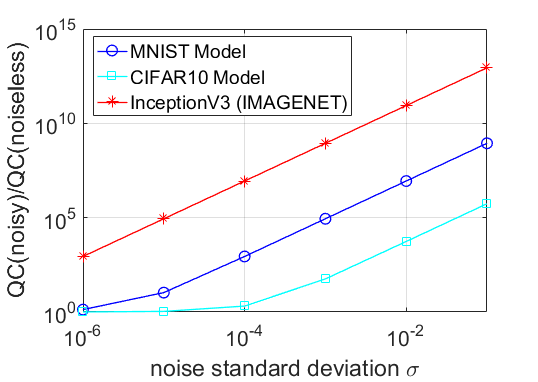}  \includegraphics[width=0.5\linewidth]{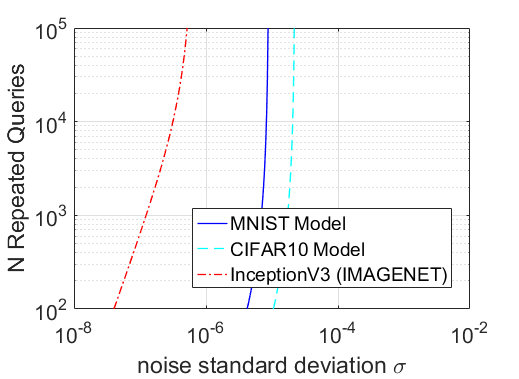}}
    
    \centerline{(a)  \hspace{0.5\linewidth} (b)} 
	\caption{(a) Ratio $R$ of query counts of noisy case to noiseless case. (b) Number of repeated queries $N$ required to estimate $a$ so that $P[\hat{a}<0]<0.3$ when $a>0$.}
	\label{fig:snr-qc}
\end{figure}

Finally, to evaluate the QC ratio $R$ with (\ref{eq1.101}), we assume $\eta = \epsilon = 0.01$, $a=0.1$, $\lambda=2$, $v_0=1$. The increase of $R$ as function of $\sigma$ is shown in Fig. \ref{fig:snr-qc}(a). At $\sigma=0.01$, we have $R=9\times 10^6, 5\times 10^3, 9\times 10^{10}$ for the three models, respectively. Considering that today's state of the art attack methods need around $10^3$ queries to attack MNIST/CIFAR10 images and $10^5$ queries to attack IMAGENET images, small noise perturbation with $\sigma=0.01$ would increase the number of queries to $10^6$ to $10^{15}$, prohibitively high to attackers. Note that the much smaller median $\Delta F_t$ values shown in Table \ref{tbl:valparams} will lead to even higher QCs. 

From Fig. \ref{fig:snr-qc}(a), for attackers with 1 million query budget, 
the defender can add noise with $\sigma=0.001, 0.01$, and $0.0001$ to mitigate them over the MNIST, CIFAR10 and IMAGENET datasets, respectively. Smaller noise, such as $\sigma \leq 10^{-4}$, is effective for well-trained models (such as MNIST) or models with a large number of classes (such as IMAGENET) that have very small $\Delta F_t$. The defender can conveniently apply appropriate noise according to its output parameters and required security level.

\begin{table}[t]
	\centering
	\begin{tabular}{c|cc|cccc}
		\toprule
		Model & ACC & Mean $F_t({\bf x})$ & $\beta$ &  Mean $\Delta F_t$ & Median $\Delta F_t$ & std $\Delta F_t$ \\
		\midrule
		MNIST Model & 0.99 & 4e-4 & 1e-3 & 5e-6 & 1e-21 & 1e-4  \\
		CIFAR10 Model & 0.83 & 1e-3 & 1e-3 & 2e-4  & 3e-8 & 1e-3  \\
		InceptionV3 & 0.78 & 3e-4 & 1e-5 & 5e-8 & 4e-9 & 1e-6 \\
		\bottomrule
	\end{tabular}
	\caption{Statistical DNN output parameters obtained from validation datasets (without noise perturbation). ACC: classification accuracy. Mean $F_t({\bf x})$: average softmax output values (excluding the top-1 pick). Mean/median/std $\Delta F_t$: mean, median, and standard deviation of output variation $\Delta F_t$.}
	\label{tbl:valparams}
\end{table}

Now we can summarize the reasons for the noise perturbation method being effective. First, from Lemma 1, the SNR of the estimated gradient becomes very low since the noise power $\sigma^2$ is amplified by the small $\beta$ and small $\Delta F_t$. Second, from Theorem 1, the gradient becomes so random that it changes the search direction to the opposite with high probability, which prevents gradient search from converge.  Finally, according to Theorem 2, low SNR makes the attack QC prohibitively high.





\subsection{Robustness to Attacker's Countermeasures} \label{counter}

The output noise perturbation method is robust to various counter-defense techniques that the attacker may adopt.
First, the attacker may try to increase $\Delta F_t$ and $\beta$ to reduce their noise amplification effects. However, $\Delta F_t$ is usually out of the attacker's control. Large $\beta$ leads to worsening gradient estimation accuracy, which in fact reduces SNR of $A$.

Second, the attacker may adopt the EOT or gradient averaging strategy that has been shown effective to invalidate gradient obfuscation defenses in white-box attack scenarios \cite{athalye2018obfuscated}. Nevertheless, EOT is not as effective in our case as one would expect. In principle, EOT finds the average gradient $\bar{\bf g} =1/J \sum_j {\bf g}_j$,  similar to (\ref{eq1.30}). Transformed images that the attacker uses to query the DNN can be written as ${\bf x}+\Delta {\bf x}_j$, where $\Delta {\bf x}_j$ denotes the difference caused by the random transform. The attacker still gets a noise perturbed output $F({\bf x}+\Delta {\bf x}_j) + {\bf v}_j$ to construct ${\bf g}_j$. The estimated gradient is still random which is worse because of the randomized $\Delta {\bf x}_j$. In this scenario, the accuracy of $\bar{\bf g}$ can not be guaranteed, even if $J$ is large. In the worst case, independent ${\bf g}_j$ may make $\bar{\bf g} \rightarrow {\bf 0}$. 

Finally, the best countermeasure is perhaps to estimate $a$ by querying the DNN repeatedly with the same ${\bf x}$. This is the optimal strategy to estimate a constant from noisy samples. Note that this is different from finding average $\bar{\bf g}$ with different ${\bf x}$.
Theoretically the attacker can average out noise and get a reliable estimation with a large number of repeated queries. We are interested to study the QC in this case.

\begin{theorem}
If the attacker conducts $N$ repeated queries with the same data ${\bf x}$, it gets $N$ samples $y_n =a+1/\beta \log z_n$, $n=1, \cdots, N$. Assume $a>0$. The minimum number of samples $N$ required to estimate $a$ as $\hat{a}$ with $P[\hat{a}<0]<\epsilon$ for some $\epsilon$ is
\begin{equation}
  N = \frac{2\sigma^2}{F_t^2({\bf x})} \left( \frac{\Phi^{-1}(\epsilon)} {e^{\frac{2\sigma^2}{F_t^2({\bf x})} - a\beta} -1} \right)^2,
 \label{eq2.10}
\end{equation}
where $\Phi^{-1}(\epsilon)$ is the inverse of the standard normal cumulative distribution function.
\end{theorem}


The detailed proof is shown in Appendix \ref{Theorem3proof}. We can see that small $F_t({\bf x})$ leads to large $N$. As a numerical example, adopting mean $F_t({\bf x})$ and $\beta$ data in Table \ref{tbl:valparams}, $\epsilon = 0.3$ and $a=1$,  the $N$ values as function of $\sigma$ are shown in Fig. \ref{fig:snr-qc}(b). A huge number of repeated queries was needed to estimate each gradient ${\bf g}_j$, which made this countermeasure technique impractical. Especially, when $\sigma \geq 10^{-4}$, no realistic $N$ could be found to estimate the gradient ${\bf g}_j$ to the correct direction with $70\%$ probability. 

\subsection{Properties of Output Noise Perturbation} \label{generalapp}

In this subsection, we first show that our analysis framework and the noise perturbation method are general enough for many other black-box attack methods. For this we present the analysis results over a ZOO-based attack \cite{tu2019autozoom}.  Then, we show that quantization noise and output-correlated noise are not effective, which explains why the output noise perturbation method is better than other randomization or gradient obfuscation methods.


Within the black-box attack community or the gradient-free optimization community, the NES and ZOO are two major gradient estimation approaches. For ZOO, we consider the AutoZOOM algorithm \cite{tu2019autozoom} that minimizes the C\&W loss function $f(t)=\log \left(F_{\max}({\bf x})/F_t({\bf x})\right)$, where  $F_{\max}({\bf x}) = \{F_i({\bf x}): i=\arg\max_j F_j({\bf x}), \forall j \neq t\}$, with the gradient estimator ${\bf g}_j = \beta^{-1} (f({\bf x}+\beta {\bf u}_j)-f({\bf x})) {\bf u}_j = a {\bf u}_j$.

\begin{theorem}
Under white Gaussian noise perturbation, the multiplication factor $a$ becomes the noisy factor $A = a+\frac{1}{\beta}\log (Z_1 Z_2)$, where $Z_1 \sim {\cal N}(1, \sigma^2/F_{\max}^2({\bf x})+\sigma^2/F_{\max}^2({\bf x}+\beta {\bf u}_j))$ and $Z_2 \sim {\cal N}(1, \sigma^2/F_t^2({\bf x}) + \sigma^2/F_t^2({\bf x}+\beta {\bf u}_j))$. In addition, when $\sigma$ is small, we have
\begin{align}
    A\sim {\cal N} \left(a,  \frac{\sigma^2}{\beta^2} \left( \frac{1}{F_{\max}^{2}({\bf x})}+ \frac{1}{F_{\max}^{2}({\bf x}+\beta {\bf u}_j)}
     +\frac{1}{F_t^{2}({\bf x})} + \frac{1}{F_t^{2}({\bf x}+\beta {\bf u}_j)} \right)\right)  \label{eq1.109}
\end{align} 
and the SNR of $A$ satisfies ${\rm SNR} \leq \frac{L^2 \beta^2}{2 \sigma^2}$.
\end{theorem}

The detailed proof is shown in Appendix \ref{Theorem4proof}.
This theorem tells us that noise perturbation randomizes the AutoZOOM's gradient estimation similarly as it does for NES.   




It can be readily seen that our analysis method shown in Theorem 1 to Theorem 4 can be applied to analyze other black-box attacks as well. For example, the ${\cal N}$attack algorithm \cite{li2019nattack} uses the NES-estimated gradients to learn adversarial distributions. Assume the adversarial samples have a certain distribution with mean $\mu$, then the ${\cal N}$attack algorithm finds $\mu$ via optimization $\mu_{t+1} = \mu_{t}-\eta \bar{\bf g}$. Obviously, the noise perturbed $\bar{\bf g}$ can hardly make the updating converge.  As another example, for the partial-information setting of \cite{ilyas2018black}, the authors propose to start from a target image. The NES algorithm is then applied to estimate the gradient to modify the target image to become similar to the original image. Noise perturbation is still effective to randomize the estimated gradients. Untargeted attacks can be analyzed similarly with just a change of loss functions.  

An especially interesting case is the hard-label attack. The label-only NES attack \cite{ilyas2018black} starts from a target image and uses its random variations to query the DNN. The binary query results are used to construct a measure similar to $F_t({\bf x})$. Obviously, noise perturbation can change the hard-label which makes this measure very noisy and reduces the SNR of the estimated gradients. Our analysis framework can still be applied.


Next, an interesting question is whether output quantization noise can be used. Another interesting question is whether the noise must be white. 

\noindent{\bf Lemma 2}. {\it Noises created by output quantization (to 2 or more bits) or noises highly correlated with DNN outputs make the random variable $Z$ have very small $\sigma_z^2$, and thus are not effective to mitigate black-box attacks.}

The detailed proof is shown in Appendix \ref{noisetypeproof}. The lemma gives a good explanation for the limited or failed defending performance of existing network randomization defenses. For example, Liu et al. \cite{liu2018towards} suggest adding noise to each convolutional layer but not the final output layer, whose net effect is to create output perturbations that are highly correlated with the true output logits. Its noise perturbation effect is in fact reduced by the network. The reduced randomization makes it more susceptible to the attacker's countermeasures such as EOT according to Fig. \ref{fig:snr-qc}(b).

\section{Experiments}
\label{experiment}

From Section \ref{formulation}, the QC needed for generating an adversarial image under our noise perturbation defense can be $10^{15}$ or more, which is computationally prohibitive. Therefore, instead of looking for QC, we followed the common practice of looking for ASR under a pre-set realistic QC limit. To save the space only ASR of targeted attacks are reported here. Experiment data for QC, distortion and untargeted attacks are presented in Appendix \ref{extraresults}. 

We experimented with a list of state-of-the-art black-box attack algorithms, including both soft-label attacks and hard-label attacks. For fair comparison, we used the original source code of the attack algorithms with their default hyper-parameters settings (represented as no-noise results). Then we inserted our noise addition defense subroutine to the source code. In practice, we could not add truly i.i.d. noise. 
Since the DNN should have softmax outputs in $[0, 1]$, we replaced negative elements with their absolute values and clipped the values over $1$. To avoid ACC degradation at high noise level (such as $\sigma=0.1$), we made the top-1 pick in the noiseless case still the top-1 pick after noise perturbation. Specifically, if the original maximum element was no longer the maximum after the noise addition, we repeatedly added absolute Gaussian noise generated by half variance to it until the original maximum element becomes maximum. Nevertheless, this was not conducted for hard-label attacks. The ACC of the noise perturbed DNN is not shown because the model accuracy degrades very little as shown in Fig. \ref{fig:snr}(a). All experiments were conducted in a machine with a single GPU.

\begin{table}[htbp]
	\centering
	\begin{tabular}{c|ccccccccc}
		\toprule
		Dataset     & Attack & QC & No &    & \multicolumn{4}{c}{Noise Standard Deviation $\sigma$} &  \\ \cmidrule{5-10}
		      &   Method & Limit & Noise & $10^{-6}$ & $10^{-5}$ & $10^{-4}$ & $10^{-3}$ & $10^{-2}$ & $10^{-1}$ \\
		\midrule
			   & ZOO & 1e5 &99.44 & 77.78 & 58.00 & 28.67 & 10.67 & 10.11 & 10.33 \\
			   & ZOO+AE & 1e5 &99.64 & 47.44 & 31.56 & 19.78 & 16.44 & 13.44 & 12.11\\
		 MNIST   
		      & AZ+AE & 1e5 &100.00 & 73.33 & 58.89 & 37.33 & 20.33 & 15.22 & 12.89 \\
		      & AZ+Bi & 1e5 & 99.89 & 91.67 & 24.78 & 17.00 & 14.00 & 12.44 & 13.44\\
		 & GenAttack & 1e5 & 95.38 & 38.19 & 30.85 & 23.61 & 12.75 & 9.85 & 5.13 \\
		 \midrule

		     &  ZOO & 1e5 &97.00 & 20.67 & 14.22 & 11.67 & 13.56 & 10.00 & 8.56 \\
		     & ZOO+AE & 1e5 & 99.00 & 52.33 & 42.00 & 32.67 & 23.44 & 17.00 & 17.22\\
 CIFAR10  
		     & AZ & 1e5 &100.00 & 70.44 & 56.78 & 42.11 & 31.22 & 19.56 & 16.67\\
		      & AZ+Bi & 1e5 & 99.33 &38.00 & 17.56 & 14.22 & 12.67 & 13.00 & 12.56\\
		 & GenAttack & 1e5 & 98.76 & 34.75 & 29.65 & 21.96 & 16.13 & 10.42 & 6.46\\
		 & Simba-pixel & 2e4 &96.31 & 96.29 & 90.93 & 43.5 & 21.23 & 12.16 & 6.9\\
		 
		 & Simba-DCT & 2e4 &97.14 & 97.14 & 89.77 & 49.85 & 27.12 & 16.39 & 10.21\\
		      		 \midrule
  &  ZOO & 2e5 & 76.00  & 0.00 & 0.00 & 0.00 & 0.00 & 0.00 & 0.00\\
		  IMAGE  & ZOO+AE & 2e5 & 92.00 & 10.00 & 0.00 & 0.00 &0.00 &0.00 &0.00 \\
		  -NET    & AZ+AE & 1e5 &100.00 & 0.00 & 0.00 & 0.00 & 0.00 & 0.00 & 0.00\\
		     & AZ+Bi & 1e5 &100.00 & 0.00 & 0.00 & 0.00 & 0.00 & 0.00 & 0.00\\
		   & NES & 1e6 & 100.00 & 80.00 & 58.00 & 20.00 & 12.00 & 2.00 & 0.00 \\
		   & GenAttack & 1e6 & 100.00 & 70.00 & 20.00 & 0.00 & 0.00 & 0.00 & 0.00\\
		   & SimBA-pixel  & 6e4 & 100.00 & 92.00 & 62.00 & 2.16 & 0.00 & 0.00 & 0.00\\
		   & SimBA-DCT & 6e4 & 96.5 & 55.00 & 50.00 & 2.13 & 0.05 & 0.00 & 0.00\\
		   
		   \cmidrule{2-10} 
		   & NES/PI & 1e6 & 93.6  & 89.4 & 21.1 & 0.00 & 0.00 & 0.00 & 0.00 \\
		
		\bottomrule
	\end{tabular}
	\caption{Targeted soft-label attacks: Attack success rate (ASR\%) versus defense noise standard deviation $\sigma$. ZOO \cite{chen2017zoo}, ZOO+AE and AZ+AE and AZ+Bi (AutoZOOM) \cite{tu2019autozoom}, GenAttack \cite{alzantot2019genattack}, SimBA-pixel and SimBA-DCT \cite{guo2019simple}, NES \cite{ilyas2018black}. NES/PI is the NES Partial Information attack where only the top-1 pick's confidence score is available \cite{ilyas2018black}. QC limit is the maximum number of iterations the attack algorithms run.}
	\label{tbl:ASRall}
\end{table}

\paragraph{ASR of Targeted Attacks:}
Table \ref{tbl:ASRall} shows the ASR of the soft-label attack methods under our proposed noise perturbation defense method. Compared with near 100\% ASR of the original noiseless case, the ASR for all the attack methods reduced significantly in presence of the proposed defense. On MNIST, noise with standard deviation $\sigma \geq 0.001$ was enough to degrade ASR from 100\% to below 20\%. On CIFAR10, noise with $\sigma \geq 0.01$ was enough. Note that ASR can not be smaller than 10\% theoretically for these two datasets because a random guess among 10 classes will result to 10\% correctness. On IMAGENET, noise with $\sigma \geq 10^{-4}$ was enough. All these results fit well with our analysis (see the end of Section \ref{formulation}). By all means, QC on the order of $10^6$ to $10^{15}$ is needed, which is much higher than the preset QC limits in these attack algorithms. 


For hard-label attacks, experiment results in Table \ref{tbl:labelonly} showed that a low noise standard deviation $\sigma=0.001$ was effective and $\sigma=0.01$ successfully reduced ASR to below $25\%$.  
Note that the observation in Fig. \ref{fig:snr}(b), i.e., the ACC did not degrade for such small $\sigma$,  was for normal images with high enough classification confidence only. The added noise could change the top-1 labels in case the classification confidence was not high enough, which happened frequently in the mid of the attacker's optimization procedure. This prevented the attack algorithm from converging.
We should also note that in hard-label attacks, the attacker needs a large number of queries, over 2.5 million queries, to generate an adversarial image even when there is no noise perturbation. This is because the gradient is already very noisy with low SNR.

\begin{table}[htbp]
\centering
\begin{tabular}{c|cccccccc}
\toprule
Dataset & Attack & No-Noise & $10^{-6}$ & $10^{-5}$ & $10^{-4}$ & $10^{-3}$ & $10^{-2}$ & $10^{-1}$ \\
\midrule
MNIST & OPT & 100.0 & 76.0 & 41.0 & 17.0 & 7.0 & \textbf{5.0} & \textbf{13.0}  \\
        & Sign-OPT & 90.0 & 71.0 & 42.0 & 14.0 & 5.0 & \textbf{0.0} & \textbf{0.0} \\
        \midrule
CIFAR10 & OPT & 81.0 & 67.0 & 58.0 & 35.0 & 33.0 & \textbf{25.0} &         \textbf{16.0} \\
        & Sign-OPT & 81.0 & 75.0 & 49.0 & 39.0 & 7.0 & \textbf{7.0} & \textbf{4.0} \\
        \midrule
& NES/Label-Only & 90.0 & 90.0 & 90.0 & 85.0 & 45.0 &  \textbf{21.0}  & \textbf{0.0} \\
IMAGENET & OPT & 100.0 & 40.0 & 20.0 & 20.0 & 10.0 & \textbf{20.0}  & \textbf{10.0} \\
& Sign-OPT & 100.0 & 60.0 & 60.0 & 30.0 & 0.0 &  \textbf{0.0}  & \textbf{0.0} \\
\bottomrule
\end{tabular}
\caption{Targeted hard-label attacks: Attack success rate (ASR\%) of the hard-label OPT attack \cite{cheng2018query}, Sign-OPT attack \cite{cheng2019sign}, and NES/Label-Only attack \cite{ilyas2018black}.}
\label{tbl:labelonly}
\end{table}

\paragraph{Comparison with Existing Defense Methods:}
In Table \ref{tbl:DefenseCompare} we compare the output noise perturbation defense method with two existing defense algorithms: JPEG Compression \cite{guo2017countering} and Input Randomization \cite{xie2018mitigating}. As seen from the table, the proposed output noise perturbation method had the best defense result with the lowest ASR. Specifically, the  two existing methods could not mitigate NES and GenAttack attacks on IMAGENET data satisfactorily, while our output noise perturbation method could reduce the ASR to near 0\%. 

\begin{table}[htbp]
	\centering
	\begin{tabular}{c|ccccc}
		\toprule
	 Dataset &	Attack     & Without & JPEG  & Randomized   & Our Noise \\
		& Method     & Defense & Compression \cite{guo2017countering}  & Input \cite{xie2018mitigating}  & Perturbation \\
		\midrule
	CIFAR10 & 	ZOO &  97.0      &  11.2  & -  & \textbf{8.56}  \\   
  & GenAttack  &  98.76   &  88.0 & 70.0 &   \textbf{10.42} \\
 \midrule
 IMAGENET & NES  &  100.0   &  66.5 & 45.9 &   \textbf{0.0} \\
   & GenAttack  &  100.0  &  89.0 & - &   \textbf{0.0 }\\
		\bottomrule
	\end{tabular}
	\caption{ASR (\%) comparison of three defense methods: output noise perturbation method, and two input randomization methods. Soft-label targeted attack.}
	\label{tbl:DefenseCompare}
\end{table}

\paragraph{Quantization and Output-Correlated Noise:}
For quantization noise, we quantized the outputs from 32-bit to 2-, 4-, and 8-bit. For output-correlated noise, the noise was generated as ${\bf v} = \alpha F({\bf x}) + \epsilon$, where $\alpha$ was the correlation coefficient and $\epsilon \sim {\cal N}({\bf 0}, 10^{-16}{\bf I})$ was the residual noise with a very small standard deviation $10^{-8}$.
Results in Table \ref{tbl:quantization4attack} clearly show that quantization did not mitigate the attacks. There was no change in ASR between the original (32-bit float) and the quantized cases. Similarly, correlated noise could not mitigate the attacks as well. The slight reduction in ASR at $\alpha=0.001$ and $\alpha=0.1$ was solely caused by the small residual noise $\epsilon$. 

\begin{table}[htbp]
\centering
\begin{tabular}{c|cc|ccc|cc}
\toprule
Dataset & Attack & Original &   \multicolumn{3}{c}{Quantization}   &  \multicolumn{2}{c} {Correlated Noise} \\ 
     & Method & No Noise & 8-bit & 4-bit & 2-bit & $\alpha=10^{-3}$ & $\alpha=0.1$ \\
\midrule
MNIST & ZOO   & 100\%  & 100\% & 100\% & 100\%  &  -   &  -\\
   &  AZ+AE & 94.5\% & 94.5\% & 94.5\% & 94.5\% & - & -\\ 
\midrule 
CIFAR10 & ZOO & 97.00\% & 100.00\% & 100.00\% & 100.00\% & - & -\\
   & AZ+AE & 99.89 \% & 99.89\% & 99.89\% & 99.89\% & 86.78\% & 84.89\% \\
\bottomrule
\end{tabular}
\caption{Attack success rate (ASR) of the ZOO and AutoZOOM black-box attack algorithms under quantization noise and output-correlated noise. Soft-label targeted attack.}
\label{tbl:quantization4attack}
\end{table}


In Appendix \ref{extraresults}.1, we presented experiment results that demonstrated the output noise perturbation method was robust to attacker's counter-measures with increased and repeated queries. 
Extra experiment results of QC, distortion, image samples as well as untargeted attacks are presented in Appendix \ref{extraresults}. 


\section{Conclusions} \label{conclusion}

In this paper, we studied the addition of white noise to DNN's output as a defense against black-box adversarial attacks. Noisy gradient is theoretically analyzed, which shows that the added noise is drastically amplified by the small logit variation. The trade-off between the defender's noise level and the attacker's query count is analyzed mathematically. Extensive experiments verified the theoretical analysis and demonstrated that white noise perturbation can effectively mitigate black-box attacks under realistic query cost constraints. 



\bibliographystyle{nipsold} 
\bibliography{my.bib}

\newpage

\centerline{\large {\bf Supplementary Material}}
\appendix

\section{Proof of Theorem 1} \label{Theorem1Proof}

As outlined in Section \ref{formulation}, the NES targeted attack algorithm minimizes the cross-entropy loss 
\begin{equation}
    f({\bf x}) = -\log F_t({\bf x})
\end{equation}
assuming the label is hot-one coded, where $F_t({\bf x})$ is the DNN output corresponding to the target class $t$. The NES algorithm minimizes the loss iteratively via gradient descent and in each iteration the gradient is estimated as
\begin{equation}
    \bar{\bf g} = 
    \frac{1}{J}\sum_{j=1}^J 
    \frac{1}{\beta}{\bf u}_j f({\bf x}+\beta {\bf u}_j),  \label{eq1.25}
\end{equation}
where $J$ queries with random direction tensors ${\bf u}_j$ are conducted to obtain DNN output $F_t({\bf x}+\beta {\bf u}_j)$ as well as loss $f({\bf x}+\beta {\bf u}_j)$. Antithetic sampling is adopted in \cite{ilyas2018black} which changes (\ref{eq1.25}) to (\ref{eq1.30}). Antithetic sampling means that both ${\bf x} +\beta {\bf u}_j$ and ${\bf x}-\beta {\bf u}_j$ are used to query the DNN.

To study the noise perturbation effect on the estimated gradient,
it is sufficient to focus on just
\begin{align}
    {\bf g}_j = {\bf u}_j \frac{1}{\beta} \log \frac{F_{t}({\bf x}-\beta {\bf u}_j)} {F_{t}({\bf x}+\beta{\bf u}_j)}.  \label{eq1.26}
\end{align}
To simplify notation, we can write ${\bf g}_j$ as the attacker-generated tensor ${\bf u}_j$ multiplying a scalar multiplication factor $a$, i.e.,
\begin{equation}
{\bf g}_j = a {\bf u}_j, \;\; a = \frac{1}{\beta} \log {h}({\bf x}), \;
h({\bf x})=\frac{F_t({\bf x}-\beta {\bf u}_j)} {F_t({\bf x}+\beta {\bf u}_j)}.  \label{eq1.27}
\end{equation}
With white Gaussian noise ${\bf v}$ added to the DNN output $F({\bf x})$, the equation (\ref{eq1.27}) becomes
\begin{equation}
    {\bf g}_j = A {\bf u}_j,  \;\;\; A =  \frac{1}{\beta} \log \tilde{h}({\bf x})
\end{equation}
where
\begin{equation}
     \tilde{h}({\bf x}) = \frac{F_t({\bf x}-\beta {\bf u}_j)+{v}_t(j+J/2)} {F_t({\bf x}+\beta {\bf u}_j)+{v}_t(j)}.
     \label{eq1.50}
\end{equation}
The variables ${v}_t(j)$ and ${v}_t(j+J/2)$ are the noises added to the target class logits $F_t({\bf x}+\beta{\bf u}_j)$ and $F_t({\bf x}-\beta{\bf u}_j)$, respectively.
Note that in antithetic sampling, we denote the noise added to the query $F_t({\bf x}-\beta {\bf u}_j)$ as $v_t(j+J/2)$, where $j+J/2$ denotes the $(j+J/2)$th query.

 The connection between the noiseless ${h}({\bf x})$ and the noisy $\tilde{h}({\bf x})$ is  
\begin{align}
   \tilde{h}({\bf x})  =  h({\bf x}) \frac{1+ \frac{v_t(j+J/2)}{F_t({\bf x}-\beta {\bf u}_j)}} {1+ \frac{v_t(j)}{F_t({\bf x}+\beta {\bf u}_j)}} \stackrel{\triangle}{=}  {h}({\bf x}) Z,    \label{eq1.55}
\end{align}
where we use the random variable $Z$ to include all the noise terms.
As a result, we have
\begin{equation}
    A = a + \frac{1}{\beta}\log Z.   \label{eq1.40}
\end{equation}
Since $Z$ is the ratio of two independent Gaussian random variables, from \cite{diaz2013existence} we can readily see that it can be approximated as a single Gaussian random variable $Z \sim {\cal N}(1, \sigma_Z^2)$ with unit mean and variance $\sigma_Z^2$ described by (\ref{eq1.45}). 

Furthermore, let $Z = 1 + S$, where $S \sim {\cal N}(0, \sigma_Z^2)$. From the small noise Definition 1, we have that $\sigma^2$ is small enough so that $\log Z = \log (1 + S) \approx S$. Therefore, from (\ref{eq1.40}) we can get $A\sim {\cal N}(a, \sigma_Z^2/\beta^2)$. Theorem 1 is proved.


{\bf Remark 1:} To understand why the noisy ${\bf g}_j$ can prevent the NES attack, it is helpful to 
have some idea about the value distribution of $h({\bf x})$, $\log h({\bf x})$ and $Z$. Since $\beta$ is very small, we expect that $h({\bf x})$ is near $1$ due to the bounded local Lipschitz constant $L$. Then, $\log h({\bf x})$ is around $0$ and can be positive or negative. The value of $a$ can also be positive and negative, and $|a|$ is usually small. This means that the factor $a$ controls the gradient descent direction. The noise $Z$ and thus $A$ make the estimated gradient ${\bf g}_j = A {\bf u}_j$ randomized, with the gradient descent direction randomized in particular. For example, even if $a$ is positive, $A$ may become negative (see the numerical example in Remark 2). 
The random multiplication factor $A$  has an accurate probability density function
\begin{equation}
    p_A(x)=\beta e^{\beta(x-a)} \frac{1}{\sqrt{2\pi}\sigma_Z} e^{-\frac{1}{2\sigma_Z^2} e^{2\beta(x-a)}}  \label{eq1.60}
\end{equation}
according to (\ref{eq1.40}). However, (\ref{eq1.60}) is too complex to conduct our subsequent SNR and QC analysis. Therefore, we have applied a further simplification to approximate $A$ as a Gaussian random variable.

\begin{figure}[htbp]
\centerline{\includegraphics[width=0.5\columnwidth]{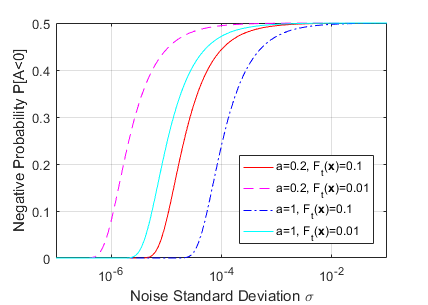}}
\caption{Probability of noisy multiplication factor $A$ becoming negative when the true value $a$ is positive. $\beta=10^{-3}$.}
\label{fig:analysis}
\end{figure}

{\bf Remark 2:}
As an example, let 
$\beta=10^{-3}$ as \cite{ilyas2018black}. For a well designed DNN, $F_t({\bf x})$ is usually around $1/C$ for a total of $C$ classes. We consider $F_t({\bf x})=0.1, 0.01$, respectively. For positive $a$, we evaluate the probability that $A$ becomes negative, which means that the gradient search direction becomes opposite to the true direction. With the cumulative distribution function (CDF) $P_A[A < x] = P_Z[Z < e^{-\beta (x-a)}]$, we can calculate $P_A[A<0]$. From the results shown in Fig. \ref{fig:analysis}, we can see that a very small noise standard deviation $\sigma = 10^{-3}$ is enough to make $P[A<0] \approx 0.5$.  




\section{Proof of Lemma 1}  \label{SNR1proof}

From (\ref{eq1.30}), i.e., the definition
\begin{equation}
    a = \frac{1}{\beta} \log \frac{F_t({\bf x}-\beta {\bf u}_j)} {F_t({\bf x}+\beta {\bf u}_j)},
\end{equation}
we have
\begin{align}
    a^2  &= \frac{1}{\beta^2} \log ^2 \left(1 + \frac{F_t({\bf x}-\beta {\bf u}_j) - F_t({\bf x}+\beta {\bf u}_j)} {F_t({\bf x}+\beta {\bf u}_j)} \right)  \nonumber \\
     &\approx  \frac{\left[ F_t({\bf x}-\beta {\bf u}_j) - F_t({\bf x}+\beta {\bf u}_j) \right]^2 }  {\beta^2 F_t^2({\bf x}+\beta {\bf u}_j)}. 
\label{eq1.62}
\end{align}
We have applied the approximation $\log(1+x)\approx x$ for small $x$ when deriving the approximation in (\ref{eq1.62}). Because $\| F_t({\bf x}-\beta {\bf u}_j) - F_t({\bf x}+\beta {\bf u}_j) \| \leq 2\beta \| {\bf u}_j \| L$, under the assumption of small $\beta$, we can guarantee $\| F_t({\bf x}-\beta {\bf u}_j) - F_t({\bf x}+\beta {\bf u}_j) \| \ll F_t({\bf x}+\beta {\bf u}_j)$ and thus the validity of (\ref{eq1.62}).
From (\ref{eq1.62}) and utilizing the approximation $\log Z = Z-1$, the SNR is then
\begin{align}
    {\rm SNR}  = \frac{a^2} {\frac{1}{\beta^2} E[(Z-1)^2]} 
    = \frac{\left[ F_t({\bf x}-\beta {\bf u}_j) - F_t({\bf x}+\beta {\bf u}_j) \right]^2 }  {F_t^2({\bf x}+\beta {\bf u}_j) \sigma_Z^2}.
\end{align}
Replacing $\sigma_Z^2$ with (\ref{eq1.30}), after some straightforward deductions we can get (\ref{eq1.100}).

Next, to derive the simplified upper bound in (\ref{eq1.100}), consider the Lipschitz constraint assumption. From the left hand side of (\ref{eq1.100}), we get
\begin{align}
  {\rm SNR} \leq
   \frac{L^2 4\beta^2 \| {\bf u}_j \|^2  F_t^2({\bf x}-\beta {\bf u}_j) }
   {\sigma^2 [F_t^2({\bf x}-\beta {\bf u}_j) + F_t^2({\bf x}+\beta {\bf u}_j)] }.
%
\end{align}
Without loss of generality, assuming $\|{\bf u}_j \|=1$ and using $F_t({\bf x}-\beta {\bf u}_j)\approx F_t({\bf x}+\beta {\bf u}_j)$, we get the SNR upper bound in the right hand side of (\ref{eq1.100}). The lemma is proved.

{\bf Remark 3:} Note that the SNR can be calculated numerically without applying the approximation in (\ref{eq1.62}). The reason we apply the approximation here is to get a simplified SNR expression that outlines the major contribution factor $\Delta F_t = |F_t({\bf x}-\beta {\bf u}_j) - F_t({\bf x}+\beta {\bf u}_j)|$.
Note also that the assumption of small $\beta$ is not a severe constraint at all in practice. In most black-box attacks, such as \cite{tu2019autozoom}, $\beta$ is selected (and proved) to be less than or equal to the inverse of DNN input dimension $d$. Obviously, $d$ is much larger than the DNN output dimension $C$ (class number). Since $F_t({\bf x}+\beta {\bf u}_j)$ on average is around $1/C$, $\beta$ is thus much less than $F_t({\bf x}+\beta {\bf u}_j)$ in most cases. This may be violated occasionally, but such occasional violations do not affect the SNR because the SNR is the average over all possible DNN outputs $F_t({\bf x})$.

\section{Proof of Theorem 2}  \label{Theorem2proof}

Consider the problem of minimizing
\begin{equation}
    f({\bf x}) = \frac{1}{2}\| F({\bf wx})-F({\bf wx}^*) \|^2
\end{equation}
with iterative gradient descent
\begin{equation}
    {\bf x}_{n+1} = {\bf x}_n - a \frac{\partial f({\bf x}_n)} {\partial {\bf x}_n},
\end{equation}
where $a$ is a constant and small learning rate. In $F({\bf wx})$, $F$ denotes the mapping of DNN, ${\bf w}$ denotes the weight of the input layer, and ${\bf x}$ denotes the input. For notation simplicity, ${\bf w}$ and ${\bf x}$ are treated as matrix and vector. ${\bf x}^*$ denotes the optimal solution. In order for the gradient descent to converge to ${\bf x}^*$ from a starting point ${\bf x}_0$ so we can count the total number of iterations, we have to assume that $F({\bf wx})$ is a monotone function between the starting point ${\bf wx}_0$ and the optimal point ${\bf wx}^*$. 

To further simplify our notation, without loss of generality, we assume $F({\bf wx})$ is a monotonously decreasing function from ${\bf wx}_0$ to ${\bf wx}^*$. We also assume ${\bf wx}_n \leq {\bf wx}^*$ for $n=0, 1, \cdots$, which can be guaranteed with a small enough learning rate $a$ and a starting point ${\bf wx}_0 \leq {\bf wx}^*$. Our subsequent deduction can be easily extended to include other cases such as $F({\bf wx})$ monotonously increasing, or some elements of $F({\bf wx})$ monotonously increasing and others decreasing, or some elements of ${\bf wx}_n$ becomes greater than ${\bf wx}^*$. In these cases, we just need to treat each element in each case individually. 

The gradient is
\begin{equation}
    \frac{\partial f({\bf x}_n)} {\partial {\bf x}_n} = {\bf w}^T F' [F({\bf wx}_n)-F({\bf wx}^*)]
\end{equation}
where $F'$ denotes the derivative of $F$ with respect to its argument ${\bf wx}_n$ and ${\bf w}^T$ denotes the transposition of ${\bf w}$. Then, the gradient updating is
\begin{equation}
    {\bf x}_{n+1} = {\bf x}_n- a {\bf w}^T F' [F({\bf wx}_n)-F({\bf wx}^*)].
\end{equation}
Next, we consider 
\begin{equation}
    {\bf wx}_{n+1} = {\bf wx}_n- a {\bf ww}^T F' [F({\bf wx}_n)-F({\bf wx}^*)]
\end{equation}
instead to exploit the assumption of ${\bf wx}_n \leq {\bf wx}^*$. 
From the Lipschitz assumption and monotonicity, we have $F({\bf wx}_n)-F({\bf wx}^*)  \leq L({\bf wx}^* - {\bf wx}_n)$ for some constant $L$. Therefore, 
\begin{equation}
   {\bf wx}_{n+1} \geq {\bf wx}_n- a {\bf ww}^T F' L ({\bf wx}^* - {\bf wx}_n).
\end{equation}
Using ${\bf wx}^*$ to subtract both sides, we get
\begin{align}
    {\bf wx}^* - {\bf wx}_{n+1} & \leq {\bf wx}^* - {\bf wx}_n  + a L{\bf ww}^T F' ({\bf wx}^* - {\bf wx}_n) \nonumber \\
    & = ({\bf I} + a L {\bf ww}^TF') ({\bf wx}^*-{\bf wx}_n) \nonumber \\
    & = ({\bf I} + a L {\bf ww}^TF')^{n+1} ({\bf wx}^*-{\bf wx}_0).  \label{eqThe2no10}
\end{align}

Denote the largest eigenvalue of the matrix $L{\bf ww}^TF'$ as $-\lambda$ where $\lambda$ is a positive value. Note that the eigenvalues must be negative because otherwise (\ref{eqThe2no10}) does not converge, which contradicts with the convergence assumption. In this special case, $F'$ is negative because $F$ is assumed monotonously decreasing.
Let $v_n = \| {\bf wx}^* - {\bf wx}_n \|$. From (\ref{eqThe2no10}), we have
\begin{equation}
    v_n \leq |1-a \lambda|^n v_0,  \label{eqA2.10}
\end{equation}
where $v_0 = \| {\bf wx}^* - {\bf wx}_0 \|$ or the initial distance from ${\bf wx}^*$. 
If $a$ is small so that $(1-a \lambda)^n \approx 1-n a\lambda$, then, in order to guarantee $v_n \leq \eta$ where $\eta$ is a small constant, the number of iterations $n$ must satisfy
\begin{equation}
   N_a \geq \frac{1-\eta/v_0}{a\lambda}.    \label{eqA2.12}
\end{equation}

Next, consider the case when the learning rate $a$ is replaced by the noisy learning rate $A = a + \sqrt{\rm SNR} v$ with noise $v \sim {\cal N}(0, 1)$. Equation (\ref{eqThe2no10}) becomes
\begin{equation}
        {\bf wx}^* - {\bf wx}_{n+1}  \leq \left(\prod_{i=0}^n ({\bf I} + A_i L {\bf ww}^TF') \right) ({\bf wx}^*-{\bf wx}_0) 
\end{equation}
where $A_i$ denotes the learning rate in the $i$th iteration. Similarly,  (\ref{eqA2.10}) becomes
\begin{equation}
    v_n \leq v_0 \prod_{i=0}^{n-1} |1- A_i \lambda|.
\end{equation}
In order to guarantee $v_n \leq \eta$, a sufficient condition is
\begin{equation}
    \prod_{i=0}^{n-1} |1-A_i\lambda| \leq \frac{\eta}{v_0}. \label{eqA2.15}
\end{equation}
If both $a$ and SNR is small, then (\ref{eqA2.15}) can be simplified to
\begin{equation}
    1 - \lambda \sum_{i=0}^{n-1} A_i \leq \frac{\eta}{v_0},
\end{equation}
which leads to
\begin{equation}
   \sum_{i=0}^{n-1} A_i \geq \frac{1-\eta/v_0} {\lambda}.
\end{equation}
Since $A_i \sim {\cal N}(a, {\rm SNR})$ are independent Gaussian random variables, in order to make  
\begin{equation}
    P\left[ \sum_{i=0}^{n-1} A_i < \frac{1-\eta/v_0}{\lambda}\right] \leq \epsilon
\end{equation}
for some small probability $\epsilon$, 
we need 
\begin{equation}
    \frac{(1-\eta/v_0)/\lambda - na} {\sqrt{n {\rm SNR}}} \leq \Phi^{-1}(\epsilon).  \label{eqA2.20}
\end{equation}
Solving (\ref{eqA2.20}) for $n$, we get that
the number of iterations needed when the learning rate becomes random $A$ must satisfy
\begin{equation}
    N_A \geq \frac{1}{4} \left[ -\frac{1}{\sqrt{\rm SNR}} \Phi^{-1}(\epsilon) + \sqrt{\frac{1}{\rm SNR} \Phi^{-2}(\epsilon) + 4 \frac{1-\eta/v_0}{a\lambda} } \right]^2.  \label{eqA2.22}
\end{equation}

Using the lower bound of (\ref{eqA2.12}) and (\ref{eqA2.22}), we can get the ratio of required iterations between the case of $a$ and the case of $A$ as
\begin{equation}
    R = \frac{N_A}{N_a} = \frac{\left[ -\frac{1}{\sqrt{\rm SNR}} \Phi^{-1}(\epsilon) + \sqrt{\frac{1}{\rm SNR} \Phi^{-2}(\epsilon) + 4 \frac{1-\eta/v_0}{a\lambda} } \right]^2} 
    {4\frac{1-\eta /v_0} {a \lambda}}
\end{equation}
which is just (\ref{eq1.101}). 
The theorem is proved.

{\bf Remark 4:} First, the proof is easier to understand if we consider ${\bf w}$ as a row vector and $F$ as a scalar nonlinear monotone function. We present the general case with the matrix ${\bf w}$ in the proof. One can actually treat each row of ${\bf w}$ separately to get the same result. Second, although $R$ is defined as the ratio of iterations, it equals to the ratio of query counts because there are a fixed number of queries conducted to estimate the gradients in each iteration. 

Third, we argue that $R$ can be used as an approximate estimation of $QC(noise)/QC(noiseless)$, i.e., the ratio of QCs between the case with noise perturbation and the case without noise perturbation in our black-box attack and defense models. It is well known that the QC expression is hard or impossible to derive for black-box attack to general DNNs because $F({\bf x})$ is highly nonlinear/nonconvex and the black-box estimated gradient is not the true gradient. The key concept of our approach is that we consider a fixed optimization trajectory of the attacker from a starting input ${\bf x}_0$ to the final adversarial input ${\bf x}^*$. This trajectory is obtained by the attack's gradient descent minimization without noise perturbation. Along this trajectory, the mapping $F({\bf x})$ can approximately be assumed as monotonously decreasing or piece-wise monotonously decreasing from ${\bf x}_0$ to ${\bf x}^*$. The attacker's estimated gradients can also be looked as true gradients with $a$ on this trajectory. The effect of noise perturbation is changing the value $a$ in each iteration to a random value $A$ with certain SNR. As a result, the model and assumptions we made for deriving $R$ in this theorem are valid for analyzing the DNN attack-defense models. Therefore, it is reasonable to claim that if the attacker uses $N_a$ iterations to get the adversarial input ${\bf x}^*$, it would needs $R$ times more iterations in case the output noise perturbation changes $a$ to $A$.  

Finally, based on the $QC(noiseless)$ needed by the attackers when there is no noise perturbation (which can be obtained by experiments), we can estimate the $QC(noise)$ needed when there is noise perturbation by multiplying $QC(noiseless)$ with $R$. By this way, we can avoid the difficulty of finding the $QC(noise)$ directly with experiments. As shown by our analysis, noise perturbation may increase $QC(noise)$ to some computationally prohibitive level, such as $10^{15}$ or more. When calculating $R$ numerically, we can use a very small $\eta/v_0$ (because $\eta$ is the desired small distance of final results ${\bf wx}_n$ to the targeting result ${\bf wx}^*$ and $v_0$ is the initial distance), and a very small $\epsilon$ (because $1-\epsilon$ is the confidence probability). We can use the average of $a$ defined in (\ref{eq1.30}) as $a\lambda$ in $R$. As a matter of fact, because SNR is usually small, the value $R$ is not very sensitive to these parameters. From (\ref{eq1.101}), it can be easily seen that $R \approx C_0/{\rm SNR}$ where $C_0$ is a small constant.

\section{Proof of Theorem 3}  \label{Theorem3proof}

Let us re-iterate the problem setting first.
In order to improve the accuracy of the estimation of ${\bf g}_j$, or specifically, the estimation of 
\begin{equation}
    a = \frac{1}{\beta} \log \frac{F_t({\bf x}-\beta {\bf u}_j)} {F_t({\bf x}+\beta {\bf u}_j)},
\end{equation}
the attacker can repeatedly query the DNN with inputs ${\bf x}-\beta {\bf u}_j$ and ${\bf x}+\beta {\bf u}_j$. The noisy outputs are $F_t({\bf x}-\beta{\bf u}_j) + v_{t1,n}$ and $F_t({\bf x}+\beta {\bf u}_j) + v_{t2,n}$ in the $n$th query, where $v_{t1,n}$ and $v_{t2,n}$ are independent Gaussian random variables ${\cal N}(0, \sigma^2)$, $n=1, \cdots, N$. 
The attacker uses the query results to calculate $y_n$ as 
\begin{equation}
    y_n = \frac{1}{\beta} \log \frac{F_t({\bf x}-\beta {\bf u}_j)+v_{t1,n}} {F_t({\bf x}+\beta {\bf u}_j)+v_{t2,n}},  \label{eq5.68}
\end{equation}
for each $n$. From Theorem 1, we have 
\begin{equation}
    y_n =a+\frac{1}{\beta} \log z_n,
\end{equation}
where $z_n \sim {\cal N}(1, \sigma_Z^2)$. To simplify notation, we let
\begin{equation}
\sigma_Z^2 = \sigma^2\left( \frac{1}{F_t^2({\bf x}-\beta {\bf u}_j)} + \frac{1}{F_t^2({\bf x}+\beta {\bf u}_j)} \right) \approx  \frac{2\sigma^2}{F_t^2({\bf x})}
\end{equation}
because $F_t({\bf x}-\beta {\bf u}_j) \approx F_t({\bf x}+\beta {\bf u}_j)$. The problem is to estimate $a$ from $y_n$, $n=1, \cdots, N$. We would like to find the $N$ that is needed for estimating $a$ reliably.

\paragraph{Lemma 3} {\it Under small noise perturbation, the optimal estimator for the attacker to estimate $a$ from $y_n$ is the maximum likelihood estimator}
\begin{equation}
    \hat{a} = \frac{1}{N} \sum_{n=1}^N y_n - \frac{1}{\beta}\sigma_Z^2.   \label{eq5.70}
\end{equation}
\paragraph{\it Proof.}  The distribution of $y_n$ is
\begin{equation}
    p_Y(y_n)= \beta e^{\beta(y_n-a)} p_Z\left(e^{\beta(y_n-a)}\right).   \label{eq5.50}
\end{equation}
From the joint distribution $p(y_1, \cdots, y_N) = \Pi_{n=1}^N p_Y(y_n)$, we can obtain the maximum likelihood estimator by making the derivative $\partial{\log p(y_1, \cdots, y_N)}/\partial{a} = 0$. With some straightforward deductions, we have
\begin{equation}
    \sum_{n=1}^N \left(e^{2\beta(y_n-a)} - e^{\beta(y_n-a)} - \sigma_Z^2\right)  = 0. \label{eq5.60}
\end{equation}
For small $\sigma$, $\log z_n$ is close to 0. Therefore, $\beta(y_n-a)$ is also very close to $0$. We can apply the approximation $e^x \approx 1 + x$ to simplify (\ref{eq5.60}) to
\begin{equation}
    \sum_{n=1}^N \left( (1+ 2\beta(y_n-a))-(1+ \beta(y_n-a)) - \sigma_Z^2\right)= 0.
\end{equation}
Then (\ref{eq5.70}) is readily available.   \hfill{$\square$} 

Now we are ready to prove Theorem 3. From (\ref{eq5.70}), after some deductions, we can get
\begin{align}
   P[\hat{a}<0]
   & = P\left[ \frac{1}{N}\sum_{n=1}^N \left(a+\frac{1}{\beta} \log z_n \right)  < \frac{\sigma_Z^2}{\beta} \right] \nonumber \\
 & = P\left[ \frac{1}{N}\sum_{n=1}^N \log z_n < \sigma_Z^2-a\beta \right].   
\end{align}
According to Jensen's inequality, $1/N\sum_n \log z_n  \leq \log 1/N\sum_n z_n$. Therefore, 
\begin{align}
    \epsilon & > P[\hat{a}<0] \nonumber \\ 
    & \geq  P\left[ \log \frac{1}{N}\sum_{n=1}^N z_n < \sigma_Z^2-a\beta \right] \nonumber \\
    & = P\left[ \frac{1}{N} \sum_{n=1}^N z_n < e^{\sigma_Z^2-a\beta} \right].  \label{eq5.80}
\end{align}
Because $1/N\sum_{n=1}^N z_n  \sim {\cal N}\left(1, \sigma_Z^2/N\right)$, (\ref{eq2.10}) can be easily found from (\ref{eq5.80}). Theorem 3 is thus proved.

{\bf Remark 5}:  There are two ways for the attacker to estimate $\hat{a}$. Besides (\ref{eq5.70}), the second way is that the attacker calculates first
\begin{align}
    \hat{F}_{1} &= \frac{1}{N} \sum_{n=1}^N \left(F_t({\bf x}-\beta {\bf u}_j) + v_{t1, n}\right),  \\
    \hat{F}_{2} &= \frac{1}{N} \sum_{n=1}^N \left(F_t({\bf x}+\beta {\bf u}_j) + v_{t2, n}\right),
\end{align}
and then estimates
\begin{equation}
 \hat{\hat{a}} = \frac{1}{\beta} \log \frac{\hat{F}_1}{\hat{F}_2}. \label{eq5.90}
\end{equation}
In this case, we have the following result:

\paragraph{Lemma 4} {\it Under small noise perturbation, we have $\hat{\hat{a}}  \approx \hat{a} \approx \frac{1}{N} \sum_{n=1}^N y_n$ if $N$ is large.}

\paragraph{\it Proof.} First, from (\ref{eq5.70}), if noise standard deviation $\sigma^2 \ll \beta F_t^2({\bf x})$, then $\hat{a} \approx \frac{1}{N} \sum_{n=1}^N y_n$. Next, starting from (\ref{eq5.90}), with the definition of $y_n$ in (\ref{eq5.68}), we have
\begin{align}
    \hat{\hat{a}} &= a + \frac{1}{\beta} \log \frac{
    \frac{1}{N}\sum_{n=1}^N \left(1 + v_{t1,n}/F_t({\bf x}-\beta {\bf u}_j) \right)}
    {\frac{1}{N}\sum_{n=1}^N \left(1+ v_{t2,n}/F_t({\bf x}+\beta {\bf u}_j) \right)}.
\end{align}
If the noise is small, we can apply the first-order Taylor series approximation $E[X/Y] \approx E[X]/E[Y]$ to get
\begin{equation}
    \hat{\hat{a}} \approx a + \frac{1}{\beta} \log \left( \frac{1}{N} \sum_{n=1}^N z_n \right). 
\end{equation}
Next, further applying first-order Taylor series approximation $\log E[X] \approx E[X]-1$, we have
\begin{align}
    \hat{\hat{a}} &\approx a + \frac{1}{\beta} \left(\frac{1}{N} \sum_{n=1}^N z_n -1 \right)   \nonumber  \\
    & = a + \frac{1}{N\beta}\sum_{n=1}^N (z_n-1) \nonumber \\
    & \approx a + \frac{1}{N\beta} \sum_{n=1}^N \log z_n  \nonumber \\
    &= \frac{1}{N} \sum_{n=1}^N \left(a + \frac{1}{\beta} \log z_n \right) \nonumber \\
    & = \frac{1}{N} \sum_{n=1}^N y_n,
\end{align}
which proves the lemma.  \hfill{$\square$}

Therefore, if $\sigma^2$ is small enough and $N$ is large enough, the estimator $\hat{a}$ in (\ref{eq5.70}) and the estimator $\hat{\hat{a}}$ in (\ref{eq5.90}) give the same result, both equal to $1/N\sum_{n=1}^N y_n$. As a result, this second way needs the similar $N$ repeated queries as (\ref{eq2.10}) in order to estimate the gradient up to certain accuracy.


\section{Proof of Theorem 4}  \label{Theorem4proof}

Consider the targeted attack toward class $t$ that is conducted by minimizing the loss function
\begin{equation}
    f({\bf x})= \log \frac{F_{\max}({\bf x})}{F_t({\bf x})},  \label{eq3.20}
\end{equation}
where $F_{\max}({\bf x}) = \{F_i({\bf x}): i=\arg\max_j F_j({\bf x}), \forall j \neq t\}$ is the logit (softmax) value of the largest non-target element, and $F_t({\bf x})$ is the logit value of the target element. Note that this is just the C\&W loss function $\max\{0, \max_{j\neq t} \log F_j({\bf x}) - \log F_t({\bf x}) \}$. The first $\max$ is skipped because we consider the adversarial search stage when the target has not been reached yet. 

If the DNN adds noise ${\bf v}\sim {\cal N}({\bf 0}, \sigma^2 {\bf I})$ to its output, then the attacker's loss becomes
\begin{equation}
    f({\bf x})= \log \frac{[F({\bf x})+{\bf v}]_{\max}}{[F({\bf x})+{\bf v}]_t},  \label{eq3.22}
\end{equation}
where $[F({\bf x})+{\bf v}]_{\max}$ and $[F({\bf x})+{\bf v}]_t$ denotes the maximum-valued element (exclude the $t$th element) and the $t$th element, respectively. 
According to the AutoZOOM attack algorithm \cite{tu2019autozoom}, the gradient estimator used by the attacker is  
\begin{align}
 {\bf g}_j &=  
    \frac{1}{\beta}{\bf u}_j \left( f({\bf x}+\beta {\bf u}_j) - f({\bf x}) \right)  \nonumber \\
    & = \frac{1}{\beta}{\bf u}_j \log \frac{\frac{[F({\bf x}+\beta {\bf u}_j)+{\bf v}_j]_{\max}}{[F({\bf x}+\beta {\bf u}_j)+{\bf v}_j]_t}} {\frac{[F({\bf x})+{\bf v}]_{\max}}{[F({\bf x})+{\bf v}]_t}}.  
    \label{eq3.25}
\end{align}
where ${\bf u}_j$ is the vector of gradient direction which is pre-set and fixed, $\beta$ is the smoothing parameter, ${\bf v}_j$ is the noise added by the DNN when the attacker queries with ${\bf x}+\beta {\bf u}_j$.

The estimated gradient equals to the vector ${\bf u}_j$ multiplying a scalar multiplication factor $A$, i.e.,
\begin{align}
{\bf g}_j = A {\bf u}_j, \;\;  A = \frac{1}{\beta} \log \tilde{h}({\bf x}), \label{eq3.29}
\end{align}
where
\begin{equation}
     \tilde{h}({\bf x}) = \frac{[F({\bf x}+\beta {\bf u}_j)+{\bf v}_j]_{\max}/[F({\bf x}+\beta {\bf u}_j)+{\bf v}_j]_t} {[F({\bf x})+{\bf v}]_{\max}/[F({\bf x})+{\bf v}]_t}.
     \label{eq3.35}
\end{equation}
Define the noiseless term
\begin{align}
  h({\bf x}) =\frac{F_{\max}({\bf x}+\beta {\bf u}_j)/F_t({\bf x}+\beta {\bf u}_j)} {F_{\max}({\bf x})/F_t({\bf x})}.  \label{eq3.30}
\end{align}
Since the noise is small, the location of the maximum element does not change almost surely. We have
\begin{align}
     \tilde{h}({\bf x})  =  h({\bf x}) \times
   \frac{1+ v_{j,\max}/F_{\max}({\bf x}+\beta {\bf u}_j)} 
   {1+v_{\max}/F_{\max}({\bf x}) }
     \times \frac{1+ v_t/F_t({\bf x})}{1+v_{j,t}/F_t({\bf x}+\beta {\bf u}_j) },    
\label{eq3.40}
\end{align}
where $v_t$ and $v_{j,t}$ are the $t$th entry of the noise vectors ${\bf v}$ and ${\bf v}_j$, respectively. The random variables $v_{\max}$ and $v_{j, \max}$ are the noises added to the maximum-valued elements of $F({\bf x})$ and $F({\bf x}+\beta {\bf u}_j)$, respectively.

Define 
\begin{align}
    Z_1 & = \frac{1+ v_{j,\max}/F_{\max}({\bf x}+\beta {\bf u}_j)} 
   {1+v_{\max}/F_{\max}({\bf x}) },  \\
   Z_2 & = \frac{1+ v_t/F_t({\bf x})}{1+v_{j,t}/F_t({\bf x}+\beta {\bf u}_j) }.
\end{align}
Each of $Z_1$ and $Z_2$ is the ratio of two independent Gaussian random variables and can be approximated as a single Gaussian random variable \cite{diaz2013existence}. Specifically, 
\begin{align}
    Z_1 &\sim {\cal N}\left(1, \;\; \frac{\sigma^2}{F_{\max}^2({\bf x})} + \frac{\sigma^2}{F_{\max}^2({\bf x}+\beta {\bf u}_j)} \right)  \\
    Z_2 &\sim {\cal N}\left(1, \;\; \frac{\sigma^2}{F_{t}^2({\bf x})} + \frac{\sigma^2}{F_{t}^2({\bf x}+\beta {\bf u}_j)} \right)  \label{eq3.45}
\end{align}
The probability density function (PDF) $p_Z(z)$ of the product $Z=Z_1 Z_2$ can be found based on \cite{cui2016exact}. 

Define $a = \frac{1}{\beta} \log h({\bf x})$. Then from (\ref{eq3.29}) and (\ref{eq3.40}) we have
\begin{equation}
     A = a + \frac{1}{\beta}\log Z_1 Z_2.  \label{eq3.47}
\end{equation}
Therefore, we can see that noise perturbation randomizes the AutoZOOM's gradient estimation similarly as it does for NES-based attack method.

To derive $A$'s distribution and SNR bound,
when noise variance $\sigma^2$ is small enough, we have $\log Z_1 \approx Z_1-1$ and $\log Z_2 \approx Z_2 -1$. Therefore, $A \approx a + \frac{1}{\beta}(Z_1-1) + \frac{1}{\beta}(Z_2-1)$, from which we can verify (\ref{eq1.109}). In addition, the SNR bound can be proved following strictly the proof of (\ref{eq1.100}). The theorem is proved.


{\bf Remark 6:} When deriving (\ref{eq3.40}), we have assumed $[F({\bf x})+{\bf v}]_{\max} = F_{\max}({\bf x})+v_{\max}$ and also $[F({\bf x}+\beta {\bf u}_j)+{\bf v}_j]_{\max} = F_{\max}({\bf x}+\beta {\bf u}_j) + v_{j,{\max}}$, which means small noise does not change the index of the maximum-valued elements. This is true almost surely under small noise perturbation. On the other hand, the noise may accidentally change the index of the maximum-valued element. In this case, the two elements, old $F_{\rm max}({\bf x})$ and new $F_{\rm newmax}({\bf x})$, have similar (almost identical) values since even tiny noise can switch their order. Therefore, (\ref{eq3.40}) is still valid.


\section{Proof of Lemma 2}  \label{noisetypeproof}

First, for output quantization, instead of outputting the full precision 32-bit logit values, the DNN can output $Q\geq 2$ bit quantized logit values. Note that 1-bit quantization is actually hard-label outputs, and only the special hard-label attack methods can work. It is well known that quantization method introduces quantization noise.  Under coarse quantization, attacks with the cross-entropy loss do not work because $a = 1/\beta \log (F_t({\bf x}-\beta {\bf u})/F_t({\bf x}+\beta {\bf u}))$ quite often results in $a=0$. However, the attacks with the C\&W loss still work well. In other words, the quantization noise can not mitigate such attacks. To explain it, let us look at the proof of Theorem 4 and consider the noise term $Z_2$. The noises are the quantization residues of $F_t({\bf x})$ and $F_t({\bf x}+\beta {\bf u}_j)$, whose quantized values are the same, say $Q$, almost surely. This means $v_t = F_t({\bf x})-Q$ and $v_{j,t} = F_t({\bf x}+\beta {\bf u}_j) - Q$. We then have
\begin{equation}
  Z_2 = \frac{2-\frac{Q}{F_t({\bf x})}}{2-\frac{Q}{F_t({\bf x}+\beta {\bf u}_j)}}.  \label{eq1.120}
\end{equation}
Obviously, $\log Z_2$ is no longer randomly positive and negative. In other words, the variance of $Z_2$ is zero.
Therefore, quantization noise can not mitigate the attacks.

Second, for output-correlated noise, let us look at the proof of Theorem 1. If the noise ${\bf v}$ is correlated to the  output $F_t({\bf x})$, then we have $v_t(j)=\sum_i \alpha_i F_t^i({\bf x} + \beta {\bf u}_j) + \epsilon$ for a very small $\epsilon \rightarrow 0$, where $\alpha_i$ are correlation coefficients. From (\ref{eq1.55}), it is easy to see that $Z$ is now randomized by $\epsilon$ only, which means a very small $\sigma_Z^2$ with much-reduced noise perturbation effect. The attack mitigation effect is also reduced.
 




\section{Extra Experiment Results}  \label{extraresults}

\paragraph{G.1 Robust to Attacker's Countermeasure with increased query counts or repeated queries:}
We evaluated the performance of the output noise perturbation method under EOT-like countermeasures, where the attackers used $N$ repeated queries to average each ${\bf g}_j$ (\ref{eq2.10}), or used more non-repeated queries (large $J$) to look for better average gradients (\ref{eq1.30}). 
First, for the countermeasure with $N$ repeated queries, from the results in Table \ref{tbl:LargeIteration}, we can say that our method was robust against this type of EOT-like countermeasures. There was no drastic change in ASR even when queries where increased to $N = 1000$, which means 3 orders of magnitude more QCs. 
Second, for the countermeasure with larger $J$ value, 
the original NES and P-RGF attack algorithms both used $J=50$. We experimented with $J=100$ and the results are summarized in Table \ref{tbl:moreJ}. From the simulation results, we can see that using a higher $J$ did not necessarily lead to better ASR. The results demonstrate that our noise perturbation method is also robust to this type of countermeasure. Note that the ASR of the untargeted attack (P-RGF) critically depends on the distortion threshold. We used the original relatively high distortion threshold for P-RGF which resulted in relatively high ASR.

\begin{table}[htbp]

	\centering
	\begin{tabular}{cccc}
		\toprule
		Dataset     & Repeated    & \multicolumn{2}{c}{Noise Standard Deviation $\sigma$}  \\
		       \cmidrule{3-4}
		      &  Queries ($N$) & $10^{-4}$  & $10^{-2}$\\
		\midrule
		     & 1   &  37.33\% & 15.22\% \\
		 MNIST    & 10  & 39.92\% & 16.73\%  \\
		  & 100 & 45.49\% & 18.33\%  \\
		      & 1000 & 48.28 \%& 22.71\% \\
		 \midrule
		     &  1  &  42.11\% & 19.56\% \\
		CIFAR10     &  10  & 45.59\%  & 23.95\%  \\
		      & 100 & 58.73\% & 27.50\% \\
		      & 1000 & 54.29\% & 29.70 \%\\
		\bottomrule
	\end{tabular}
	\caption{Attack success rate (ASR) of the AutoZOOM targeted attack \cite{tu2019autozoom} under noise perturbation defense along with the countermeasure where the attacker used $N$ repeated queries to average gradients.}
	\label{tbl:LargeIteration}
\end{table}

\begin{table}[htbp]
\centering
\begin{tabular}{cccc}
\toprule
Attack & $J=50$ & $J=100$  \\
\midrule
 NES (targeted) & 20.00\% &     15.78\%   \\
 P-RGF (untargeted) & 60.00\% & 57.89\% \\
\bottomrule
\end{tabular}
\caption{Attack success rate (ASR) of the NES \cite{ilyas2018black} and P-RGF \cite{cheng2019improving} black-box attack algorithms under noise perturbation defense with noise standard deviation $\sigma=10^{-4}$. IMAGENET. P-RGF: transfer-learning prior-guided random gradient free algorithm \cite{cheng2019improving}.}
\label{tbl:moreJ}
\end{table}



\paragraph{G.2 ASR of Untargeted Attacks:}
For untargeted attacks, we just need to change the loss function to $f({\bf x}) = \log F_t({\bf x})$
for the cross-entropy loss or $f({\bf x})=  \log \frac{F_{t}({\bf x})}{F_{\max}({\bf x})}$ for the C\&W loss, where $F_t({\bf x})$ is the true logit value of the input ${\bf x}$, and $F_{\max}({\bf x})$ denotes the maximum logit value excluding the true logit.  Our analysis and conclusions derived in Section \ref{analysis} are still valid. Specifically, we still have noisy multiplication factor $A=a + \beta^{-1} \log Z$ with an extremely low SNR which leads to high QC and low ASR. 
This is demonstrated by our experimental results in Table \ref{tbl:ASRall-untarget} for soft-label attacks and Table \ref{tbl:labelonly-untarget} for hard-label attacks. The additive noise with $\sigma = 0.01$ successfully mitigated all these attack methods.


\begin{table}[htbp]
	\centering
	\begin{tabular}{c|cccccccc}
		\toprule
	Dataset     & Attack & No &    & \multicolumn{4}{c}{Noise Standard Deviation $\sigma$} &  \\ \cmidrule{4-9}
		  &   Method & Noise & $10^{-6}$ & $10^{-5}$ & $10^{-4}$ & $10^{-3}$ & $10^{-2}$ & $10^{-1}$ \\
		\midrule
		
		 & ZOO & 100.00 & 63.99 & 60.99  & 53.99  &  17.99 & \textbf{0.99}  & \textbf{0.0} \\
		MNIST     
		      & AZ+AE &  100.00 & 22.99 & 21.00  & 13.99  & 20.03  & \textbf{0.00} & \textbf{0.00} \\	
		\midrule

		     &  ZOO & 100.00 & 47.00  & 42.00  & 33.99  & 27.00  & \textbf{26.00}  & \textbf{19.99}  \\
	  CIFAR10 
		      & AZ+AE & 100.00 & 60.00 & 53.00 & 52.00 & 39.99 & \textbf{14.00} & \textbf{10.00} \\
		      & SimBA-pixel  & 92.59 & 91.84 & 59.84 & 22.48 & 7.75 & \textbf{2.69} & \textbf{1.21} \\
		      & SimBA-DCT & 93.3 & 92.33 & 64.53 & 27.53 & 8.52 & \textbf{3.47} & \textbf{1.56}\\
		      \midrule
 &  ZOO & 86.00 & 20.00 & 10.00 & 0.00 & 0.00 & \textbf{0.00} & \textbf{0.00} \\
 
		 IMAGENET
		      & AZ+AE & 100.00 & 100.00 & 100.00 & 50.00 & 20.00 & \textbf{20.00} & \textbf{10.00} \\
		      & SimBA-pixel  & 93.69 & 93.72 & 93.53 & 76.8 & 0.00 & \textbf{0.00} & \textbf{0.00} \\
		      & SimBA-DCT  & 90.01 & 90.01 & 90.01 & 86.57 & 0.00 & \textbf{0.00} & \textbf{0.00} \\
		      & P-RGF & 98.0 & 96.00 & 90.0 & 60.0 & 46.0 &  \textbf{36.0}  & \textbf{30.0} \\
		
		\bottomrule
	\end{tabular}
	\caption{Untargeted soft-label attacks: Attack success rate (ASR\%) versus defense noise standard deviation $\sigma$. ZOO \cite{chen2017zoo}, AZ+AE \cite{tu2019autozoom}, SimBA-pixel, SimBA-DCT \cite{guo2019simple}, and P-RGF attack \cite{cheng2019improving}.}
	\label{tbl:ASRall-untarget}
\end{table}

\begin{table}[htbp]
\centering
\begin{tabular}{c|cccccccc}
\toprule
Dataset & Attack & No  &    & \multicolumn{4}{c}{Noise Standard Deviation $\sigma$} &  \\ \cmidrule{4-9}
& Method &
Noise & $10^{-6}$ & $10^{-5}$ & $10^{-4}$ & $10^{-3}$ & $10^{-2}$ & $10^{-1}$ \\
\midrule
MNIST & OPT & 98.0 & 95.0 & 86.0 & 35.0 & 22.0 & \textbf{16.0} & \textbf{9.0}  \\
        & Sign-OPT & 98.0 & 78.0 & 37.0 & 5.0 & 3.0 & \textbf{4.0} & \textbf{0.0} \\
        & Boundary & 100.00 & 100.00 & 100.00 & 100.00 & 96.00 & \textbf{64.00} &  \textbf{2.00}\\
        \midrule
CIFAR10 & OPT & 100.0 & 81.0 & 63.0 & 18.0 & 18.0 & \textbf{17.0} &         \textbf{13.0} \\
        & Sign-OPT & 100.0 & 69.88 & 59.04 & 31.33 & 2.41 & \textbf{0.0} & \textbf{0.0} \\
        & Boundary & 100.00 & 100.00 & 100.00 & 100.00 & 98.00 & \textbf{88.00} & \textbf{9.00} \\
        \midrule

IMAGENET & OPT & 100.0 & 60.0 & 60.0 & 0.0 & 0.0 & \textbf{0.0}  & \textbf{0.0} \\
& Sign-OPT & 100.0 & 50.0 & 50.0 & 30.0 & 0.0 &  \textbf{0.0}  & \textbf{0.0} \\
\bottomrule
\end{tabular}
\caption{Untargeted hard-label attacks: Attack success rate (ASR\%) of the hard-label OPT attack \cite{cheng2018query}, Sign-OPT attack \cite{cheng2019sign}, and Boundary \cite{brendel2017decision}.}
\label{tbl:labelonly-untarget}
\end{table}

{\bf Remark 7:} Before proceeding, we would like to point out that this experiment demonstrated that the output noise perturbation defense was effective against the transfer-learning based attack \cite{cheng2018query} as well as the boundary-based attack \cite{brendel2017decision}, both of which were considered as free of gradient estimation. The P-RGF attack \cite{cheng2018query} applied transfer-learning model to assist gradient estimation, while the boundary-based attack \cite{brendel2017decision} did not have explicit gradient calculation. Experimental data in Table \ref{tbl:ASRall-untarget} showed that the P-RGF had ASR reduced from 98\% to 36\% under $\sigma=0.01$. The relatively high ASR of 36\% was due to the strong transfer model and the larger $L_2$ distortion threshold used in the original source code. It is well known that untargeted attacks can be always successful as long as large distortion can be allowed. The ASR would drop to very low level if the transfer model did not fit well with the black-box DNN or a small $L_2$ distortion level was used. In contrast, the boundary attack data in Table \ref{tbl:labelonly-untarget} showed very low ASR at $\sigma=0.1$. Although the ASR was relatively high at $\sigma=0.01$, extremely high QCs were used (see Table \ref{tbl:ASRall01}). Limiting the QC realistically would reduce ASR and prevent this attack with our defense.

In Table \ref{tbl:ASRall0} we show that our noise perturbation method was effective to mitigate the P-RGF untargeted attack algorithm with various IMAGENET classification models. The ASR reduced with the increase in perturbation level. 

\begin{table}[htbp]
	\centering
	\begin{tabular}{ccccccccc}
		\toprule

		 Classification    & Attack & No &    & \multicolumn{3}{c}{Noise Standard Deviation $\sigma$} & & \\ \cmidrule{4-9}
		         Model &  Method & Noise & $10^{-6}$ & $10^{-5}$ & $10^{-4}$ & $10^{-3}$ & $10^{-2}$ & $10^{-1}$ \\
		\midrule
		   Inception-v3&  & 98.00 & 96.00 & 90.00 & 60.00 & 46.00 & 36.00 & 30.00 \\
		   VGG-16   & P-RGF & 100 & 100 & 89.8 & 71.43 & 46.94 & 48.98 & 44.9 \\
		       ResNet-50& & 97.91 & 95.83 & 89.58 & 83.33 & 75.00 & 56.25 & 50.00\\
		\bottomrule
	\end{tabular}
	\caption{Attack Success Rate (ASR\%) versus defense noise standard deviation $\sigma$ for P-RGF over IMAGENET for different pre-trained models. Untargeted attack.}
	\label{tbl:ASRall0}
\end{table}

In Table \ref{tbl:DefenseCompare2} we compare the noise perturbation defense method with two existing defense algorithms: JPEG Compression \cite{guo2017countering} and Input Randomization \cite{xie2018mitigating}. As seen from the table, the proposed noise perturbation method had the best defense result. 

\begin{table}[htbp]
	\centering
	\begin{tabular}{c|ccccc}
		\toprule
	 Dataset &	Attack     & Without & JPEG  & Randomized   & Our Noise \\
		& Method     & Defense & Compress \cite{guo2017countering}  & Input \cite{xie2018mitigating}  & Perturbation \\
		\midrule
 CIFAR10 &   ZOO  &   100.0    &  28.7   & -  &  \textbf{20.0} \\ 
  IMAGENET & P-RGF  &  98.0  & 81.1  & 82.3    & \textbf{30.0} \\
		\bottomrule
	\end{tabular}
	\caption{ASR (\%) comparison of three defense methods in untargeted attack setting.}
	\label{tbl:DefenseCompare2}
\end{table}

\paragraph{G.3 Query Count and Distortion of Targeted/Untargeted Attacks:}
While Table \ref{tbl:ASRall} shows the ASR of the attack algorithms under our noise perturbation defense, we also obtained their average query counts and $L_2$ per-pixel distortion, which are shown in 
Table \ref{tbl:ASRall01} and Table \ref{tbl:l2dist}. 
The query counts were calculated with successful adversarial images under the $L_2$ distortion threshold. The 0 query count means there were  no such successful adversarial images available. We can see that besides ASR reduction, the query efficiency of all the attack algorithms drastically reduced due to noise perturbation. In general, the query count gradually increased with the increased noise perturbation level $\sigma$. However, after a certain noise level, when the number of successful adversarial images reduced to a certain level, we would have query counts less than those of noiseless case. This might be because the remaining successful adversarial images were those that were easy to attack. 

There were some successful adversarial images depending on the level of noise perturbations. We are interested to check whether these images were truly successful attacks. A successful attack requires low enough distortion. Therefore, we checked the $L_2$ distortion of these adversarial images. The comparison of $L_2$ distortion between the noiseless attacks and noise defenses is shown in Table \ref{tbl:l2dist}. We can see that the $L_2$ distortion increased with noise level $\sigma$. The distortion under $\sigma=0.01$ was several times larger than those of noiseless attack. This means that even if the attacks were considered successful, the adversarial samples had high distortion. On the IMAGENET dataset, the ZOO and AutoZOOM algorithms had no data in targeted attacks because their ASR was 0. P-RGF, NES, AutoZOOM, and ZOO had $L_2$ distortion thresholds ranked from large to small. Their ASR also ranked from high to low under noise perturbation. 

\begin{sidewaystable}[htbp]
	\centering
	\begin{tabular}{c|c|c rrrrrrr}
		\toprule
		& Dataset     & Attack & No &    & \multicolumn{3}{c}{Noise Standard Deviation $\sigma$} & & \\ \cmidrule{5-10}
		   &   &   Method & Noise & $10^{-6}$ & $10^{-5}$ & $10^{-4}$ & $10^{-3}$ & $10^{-2}$ & $10^{-1}$ \\
		\midrule
				  & & ZOO & 38096 & 51312 & 57320 & 52049 & 3658 & 4543 & 9370 \\
	&	 MNIST    
		      & AZ+AE & 690 & 510 & 491 & 466 & 336 & 263 & 329\\

	&	      & OPT & 68704 & 68875 & 65832 & 53988 & 17383 & 13200 & 15181 \\
	&	      & Sign-OPT & 62529 & 62684 & 62317 & 56594 & 28029 & 9976 & 7544 \\
		 \cmidrule{2-10}

	&	     &  ZOO & 28096 & 34008 & 40403 & 35460 & 45793 & 28610 & 10029 \\
		     
Targeted	&	  CIFAR10  
		      & AZ+AE & 164 & 214 & 233 & 253 & 375 & 362 & 253 \\

     &           & SimBA-pixel  & 269 & 271 & 774 & 5190 & 7251 & 8089 & 8574 \\
	&	      & SimBA-DCT & 214 & 214 & 901 & 4608 & 6710 & 7699 & 8269 \\
	&	      & OPT & 69023 & 66595 & 64122 & 45997 & 18100 & 11094 & 10644 \\
	&	      & Sign-OPT & 59307 & 59550 & 59306 & 52655 & 29312 & 13825 & 10960 \\
		  \cmidrule{2-10}

	&	  &  ZOO & - & 0.0 & 0.0 & 0.0 & 0.0 & 0.0 & 0.0\\
Attacks	&	  IMAGENET
		      & AZ+AE & - & 0.0 & 0.0 & 0.0 & 0.0 & 0.0 & 0.0\\

	&	      & NES & 80577  & 64799 & 34608 & 14788 & 1214 & 2870 & 0 \\
	&	      & SimBA-pixel  & 8828 & 88024 & 89433 & 89161 & 89992 & 89992 & 89992 \\
	&	      & SimBA-DCT & 6000 & 6890 & 6990 & 7013 & 7046 & 7050 & 7050 \\
	&	      & OPT & 355856  & 327497 & 295705 & 180335 & 53130 & 17660 & 2190 \\
	&	      & Sign-OPT & 70452 & 70748 & 70548 & 69858 & 48823 & 13956 & 7775\\
		\midrule
		\midrule
		  & & ZOO & 12492 & 63584 & 68528 & 78866 & 101973 & 60928 & 0 \\
	&	 MNIST    
		      & AZ+AE & 806 & 1128 & 999 & 950 & 537 & 0 & 0\\
	&	      & OPT & 84279 & 84293 & 84283 & 77630 & 24915 & 7261 & 6548 \\
	&	      & Sign-OPT & 39730 & 39626 & 37543 & 35444 & 25373 & 9922 & 6530 \\
	&        & Boundary & 1055600 & 1055665 & 1055653 & 1055458 & 1055552 & 1055658 & 1055297 \\
		 \cmidrule{2-10}

	&	     &  ZOO &13834 & 9352 & 32499 & 38656 & 39006 & 26525 & 31987 \\
		     
Untargeted	&	  CIFAR10  
		      & AZ+AE & 535 & 534 & 650 & 672 & 646 & 733 & 990 \\
     &           & SimBA-pixel  & 555 & 624 & 3631 & 7124 & 8493 & 8961 & 9098 \\
	&	      & SimBA-DCT & 522 & 612 & 3217 & 6661 & 8423 & 8890 & 9065 \\
	&	      & OPT & 65536 & 65092 & 62688 & 45312 & 16338 & 8034 & 7701 \\
	&	      & Sign-OPT & 30437 & 30457 & 30102 & 28415 & 16964 & 7987 & 6627 \\
	&        & Boundary & 1057026 & 1057007 & 105684 & 1056919 & 1056855 & 1056179 &  1056162\\
		      		 \cmidrule{2-10}

	&	  &  ZOO & 271627 & 13586928 & 3230464 & 0 & 0 & 0 & 0\\
Attacks	&	  IMAGENET    
		      & AZ+AE & 4999 & 5568 & 15668 & 59096 & 186415 & 87779 & 63349\\
	&	      & P-RGF & 32434 & 38890 & 46864 & 58458 & 49226 & 41040 & 28798 \\
	&	      & SimBA-pixel  & 1447 & 1443 & 1519 & 6663 & 29992 & 29993 & 29993 \\ 
	&	      & SimBA-DCT & 575 & 564 & 564 & 814 & 7049 & 7050 & 7050 \\
	&	      & OPT & 76804 & 76855 & 73218 & 28540 & 4061 & 2361 & 2490 \\
	&	      & Sign-OPT & 65320 & 65341 & 65393 & 40370 & 15509 & 7358 & 5964 \\
		
		\bottomrule
	\end{tabular}
	\caption{Query count versus defense noise standard deviation $\sigma$. Query count is the average query count for those successfully attack image samples.}
	\label{tbl:ASRall01}
\end{sidewaystable}

\centering
\begin{table}[htbp]
  \centering
	\begin{tabular}{c|c|cc ccc ccc}
		\toprule
	&	Dataset     & Attack  & No & \multicolumn{6}{c}{Noise Standard Deviation $\sigma$} \\ \cmidrule{5-10}
		 &    & Method   &  Noise & $10^{-6}$ & $10^{-5}$ & $10^{-4}$ & $10^{-3}$ & $10^{-2}$ & $10^{-1}$ \\
		\midrule
		
		&	 & ZOO  &  29.27 & 79.34& 81.12 & 95.56 & 111.7 &112.32& 115.28\\
	&	MNIST     
		      & AZ+AE &  77.17 & 82.79 & 95.17 & 109.6 & 143.77 & 163.73 &  167.87\\
		      & & OPT & 1.85 & 1.78 & 1.79 & 1.93 & 2.39 & 3.86 & 4.05 \\
	        & & Sign-OPT & 1.54 & 1.25 & 0.88 & 0.29 & 0.25 & 3.79 & 0.27 \\
		 \cmidrule{2-10}

		&     &  ZOO  &  3.54 &  13.74 & 23.54 & 34.07 & 41.59 & 48.19 &  57.27\\
Targeted	&	  CIFAR10  
		      & AZ+AE &  19.63 & 22.22 & 24.8 & 28.87 & 35.13 & 62.03 & 68.71\\
		      & & SimBA-pixel & 1.48 & 1.57 & 8.59 & 7.56 & 5.8 & 5.3 & 5.18 \\ 
	          & & SimBA-DCT & 1.23 & 1.22 & 9.61 & 7.53 & 5.75 & 5.26 & 5.12 \\
	       & & OPT & 0.37 & 0.42 & 0.39 & 0.5 & 1.17 & 2.83 & 3.3 \\
	        & & Sign-OPT & 0.25 & 0.23 & 0.19 & 0.15 & 0.19 & 0.3 & 0.31 \\
		 \cmidrule{2-10}

Attacks		& & ZOO & - & - & - & - & - & - & -  \\
	 & IMAGENET 
	      & AZ+AE &- & - & - & - & - & - & -\\
	      &	 & NES & 0.41  & 2.99 & 4.3 & 4.22 & 11.18 & 18.46 & 21.97\\
	      & & SimBA-pixel & 14.22 & 7.22 & 6.3 & - & - & - & - \\ 
	          & & SimBA-DCT & 12.22 & 6.02 & 5.35 & 5.25 & 5.09 & - & - \\
	       & & OPT & 53.37 & 58.19 & 69.52 & 68.58 & 109.69 & 126.28 & 129.11 \\
	        & & Sign-OPT & 5.34 & 5.12 & 5.48 & 7.62 & - & - & - \\
	      
	     \midrule \midrule
	     
	     &		 & ZOO  & 20.36& 33.08 & 34.03 & 36.54 & 52.3 & 66.81 & 74.9 \\
	&	MNIST      
		      & AZ+AE & 35.8 & 49.2 & 52.89 & 56.46 & 61.35 & 71.74 & 73.31\\
		      & & OPT & 1.07 & 1.07 & 1.07 & 1.1 & 1.94 & 5.12 & 5.52 \\
	        & & Sign-OPT & 1.03 & 1.04 & 1.04 & 1.04 & 1.26 & 1.25 & 5.37\\
	        &        & Boundary & 61.93 & 61.49 & 61.15 & 61.41 & 61.14 & 61.41 & 74.35 \\
		    \cmidrule{2-10}
	     
		   &  &  ZOO  & 2.4 & 34.5 & 32.4 & 54.4 & 61.8 & 73.2 & 84.4\\
Untargeted &		  CIFAR10    
		      & AZ+AE & 13.6 & 21.5 & 14.16 & 20.52 & 23.26 & 21.92 & 35.92\\
		      & & SimBA-pixel & 3.81 & 4.94 & 14.11 & 7.43 & 5.58 & 5.28 & 5.14 \\ 
	          & & SimBA-DCT & 3.579 & 5.36 & 13.68 & 7.61 & 5.54 & 5.28 & 5.16 \\
	       & & OPT & 0.15 & 0.15 & 0.15 & 0.18 & 0.6 & 2.03 & 2.5 \\
	        & & Sign-OPT & 0.12 & 0.12 & 0.12 & 0.13 & 0.3 & 1.36 & 2.51\\
	        &        & Boundary & 28.68 & 28.57 & 28.62 & 28.63 & 29.12 & 29.05 & 29.69 \\
		      
		 \cmidrule{2-10}
Attacks &		 & ZOO & 0.0023 & 21.8 & 23.3 & 25.4 & 25.4 & 25.35 & 27.00  \\
	& IMAGENET 
	      & AZ+AE & 0.27 & 0.27 & 0.26 & 1.02 & 1.65 & 2.07 & 2.37\\
	   &  & P-RGF & 8.34 & 6.92 & 8.89 & 23.63 & 29.3 & 34.01 & 37.55 \\
	   & & SimBA-pixel & 0.19 & 0.19 & 0.19 & 1.02 & - & - & - \\
	   & & SimBA-DCT & 0.19 & 0.2 & 0.21 & 0.32 & - & - & - \\ 
	   & & OPT & 21.54 & 21.53 & 22.65 & - & - & - & - \\
	   & & Sign-OPT & 1.70 & 8.52 & 9.68 & 24.47 & - & - & - \\
	      
		\bottomrule
	\end{tabular}
    \caption{Per-pixel $L_2$ distortion ($\times 10^{-4}$) versus noise perturbation standard deviation $\sigma$.}
  \label{tbl:l2dist}
\end{table}

\paragraph{G.4 Sample Adversarial Images:}

\begin{figure}[htbp]
\centering
\fbox{\includegraphics[width=0.95\linewidth]{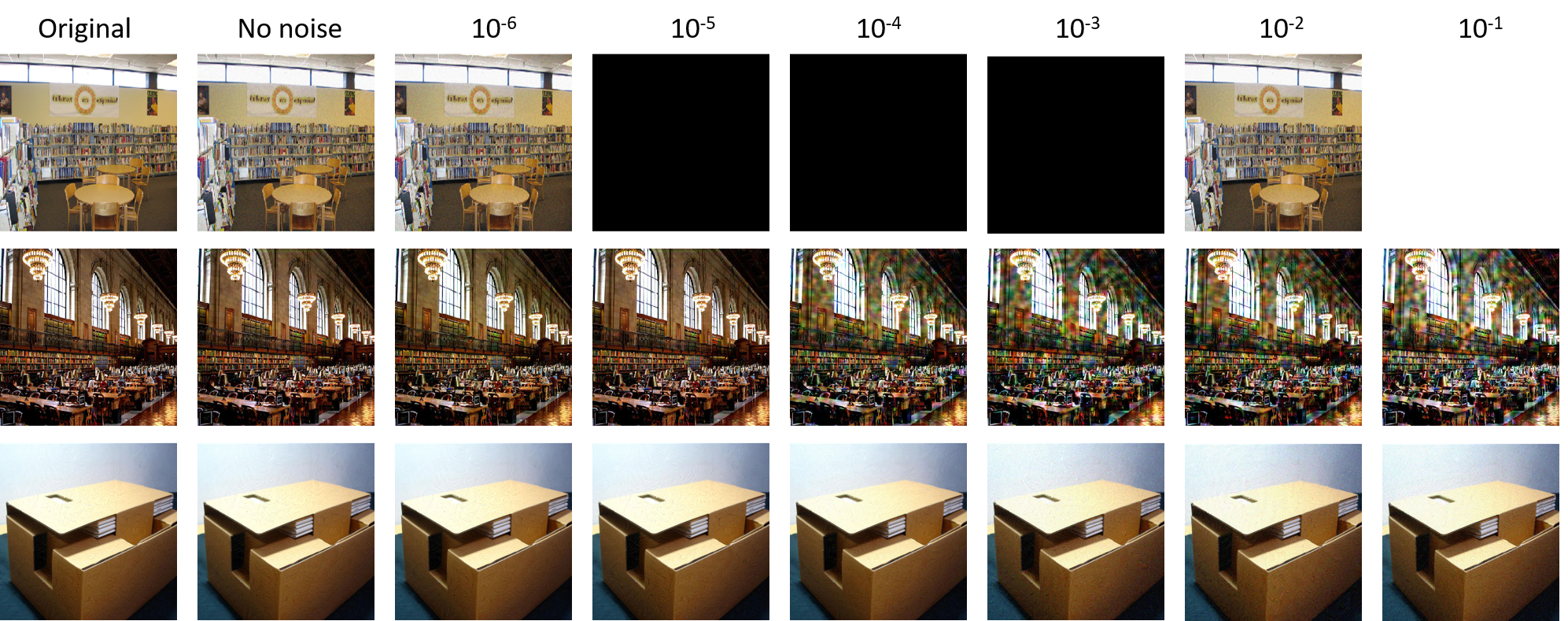}}
\caption{IMAGENET successful adversarial samples. From top to bottom: Samples obtained by NES (targeted), AutoZOOM (untargeted), P-RGF (untargeted) attack algorithms, respectively. From left to right: Original images and adversarial images obtained at $\sigma=0, 10^{-6}, 10^{-5}, 10^{-4}, 10^{-3}, 10^{-2},10^{-1}$, respectively.}
\label{fig:imagesample10}
\end{figure}

\begin{figure}[t]
\centering
\fbox{\includegraphics[width=0.75\columnwidth]{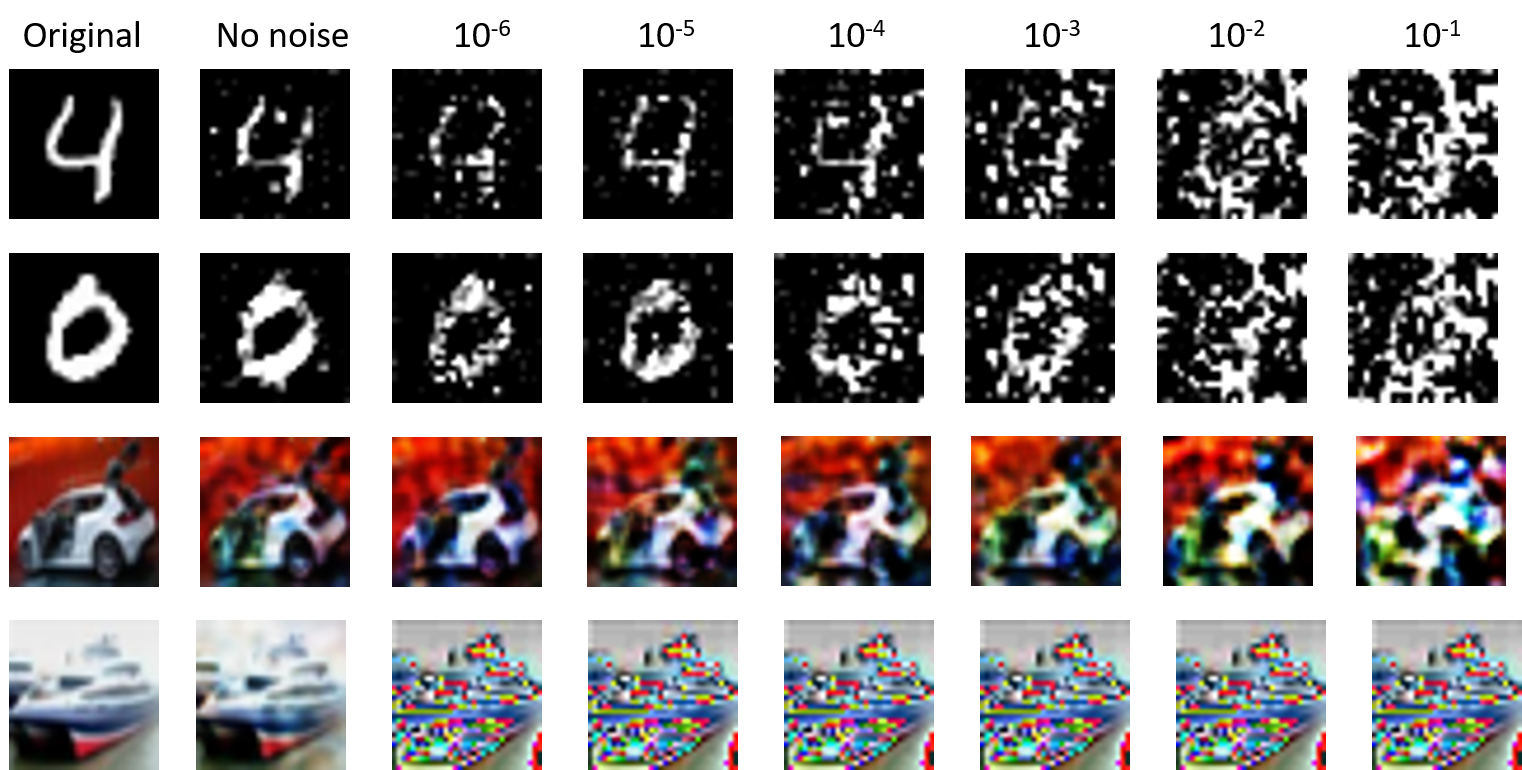}}
\fbox{\includegraphics[width=0.75\columnwidth]{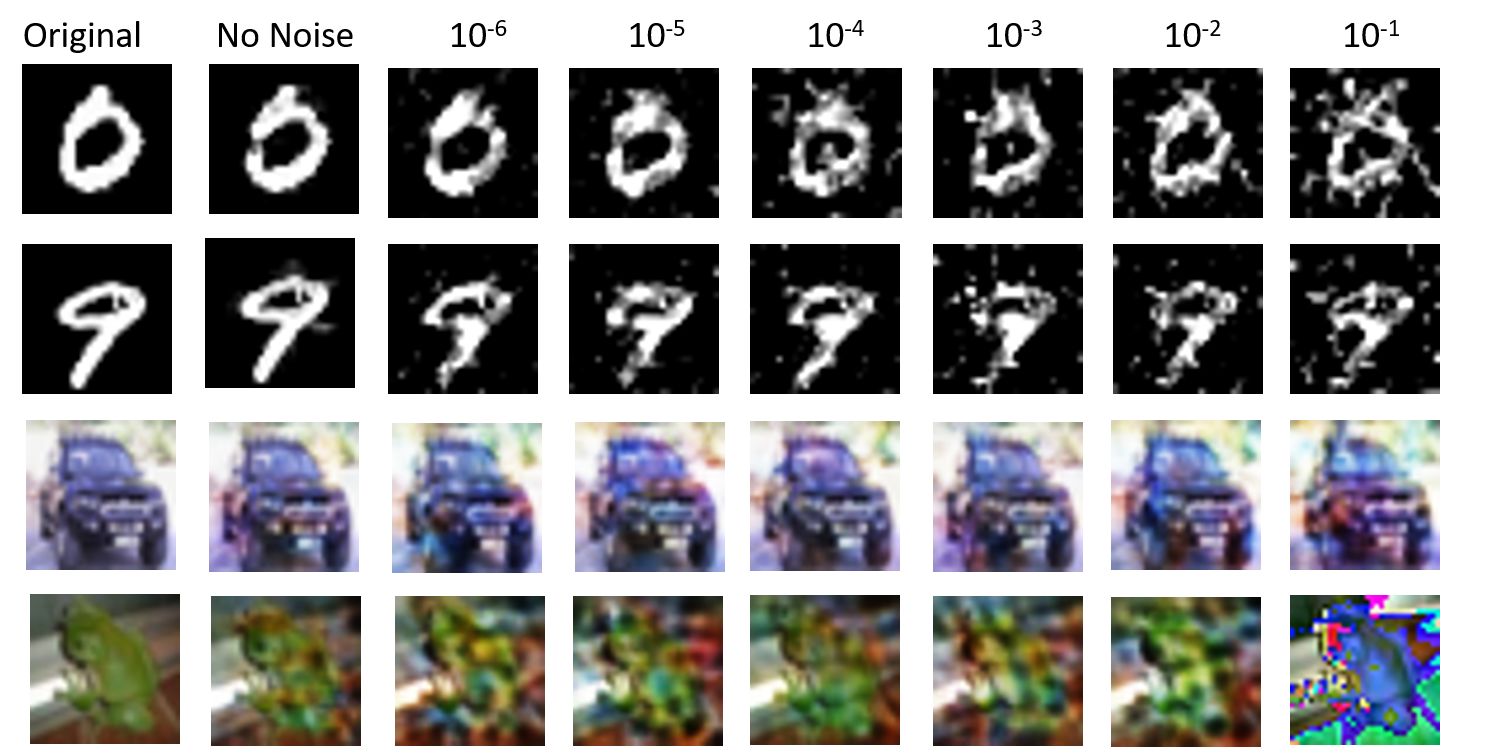}}
\caption{MNIST and CIFAR10 successful adversarial samples obtained by AutoZOOM. Top: targeted attack. Bottom: untargeted attack. From left to right: Original images, Adversarial outputs at $\sigma=0, 10^{-6}, 10^{-5}, 10^{-4}, 10^{-3}, 10^{-2}, 10^{-1}$.}
\label{fig:imagesample}
\end{figure}

Fig. \ref{fig:imagesample10} shows some sample IMAGENET images generated by the adversarial algorithms. Heavier distortions can be seen when $\sigma \geq 10^{-4}$. Especially, some images obtained by the NES targeted attacks were black-out but were still classified as successful attacks. 
Fig. \ref{fig:imagesample} shows the visual effects of adversarial MNIST and CIFAR10 images. Similarly, the adversarial examples at higher $\sigma$ could no longer deceive human perception, especially in targeted attacks. 

%


\end{document}